\documentclass[aps,prl,twocolumn,showpacs,preprintnumbers,amsmath,amssymb]{revtex4-1}
\usepackage{graphicx}
\usepackage[usenames,dvipsnames]{xcolor}
\usepackage{hyperref}
\hypersetup{colorlinks=true, citecolor=blue, urlcolor=blue, linkcolor=blue, unicode=true}
\usepackage[version=4]{mhchem}
\usepackage{cleveref}
\usepackage{makecell}
\usepackage{float}
\usepackage{lineno}
\usepackage[T1]{fontenc}
\usepackage[caption=false]{subfig}
\newcommand{\Tr}{\textrm{Tr}}

\newcommand{\bra}[1]{\ensuremath{\langle #1 |}}
\newcommand{\ket}[1]{\ensuremath{| #1 \rangle}}

\usepackage{amssymb,amsmath} 
\usepackage{amsfonts}
\usepackage{color}
\usepackage[utf8]{inputenc}
\usepackage{lmodern}
\usepackage{epsfig}
\usepackage{ulem}
\usepackage{blindtext}
\begin{document}
\title{Probing quantum floating phases in Rydberg atom arrays}
\author{Jin Zhang$^{1,2}$}
\thanks{These authors contributed equally to this work}
\email{jzhang91@cqu.edu.cn}
\author{Sergio H. Cant\'u$^{3}$ }
\thanks{These authors contributed equally to this work}
\email{scantu@quera.com}
\author{Fangli Liu$^{3}$}
\email{fliu@quera.com}
\author{Alexei Bylinskii$^{3}$}
\author{Boris Braverman$^{3}$}
\author{Florian Huber$^{3}$}
\author{Jesse Amato-Grill$^{3}$}
\author{Alexander Lukin$^{3}$}
\author{Nathan Gemelke$^{3}$}
\author{Alexander Keesling$^{3}$}
\author{Sheng-Tao Wang$^{3}$}
\author{Y. Meurice$^1$}
\author{S.-W. Tsai$^4$}
\affiliation{$^1$Department of Physics and Astronomy, University of Iowa, Iowa City, IA 52242, USA}
\affiliation{$^2$ Department of Physics and Chongqing Key Laboratory for Strongly Coupled Physics, Chongqing University, Chongqing 401331, China}
\affiliation{$^3$QuEra Computing Inc., 1284 Soldiers Field Road, Boston, MA, 02135, USA}
\affiliation{$^4$Department of Physics and Astronomy, University of California, Riverside, CA 92521, USA}

\definecolor{burnt}{cmyk}{0.2,0.8,1,0}
\def\lt{\lambda ^t}
\def\note{note}
\def\beq{\begin{equation}}
\def\enq{\end{equation}}
\def\ya{\textcolor{red}}
\date{\today}
\begin{abstract}
The floating phase, a critical incommensurate phase, has been theoretically predicted as a potential intermediate phase between crystalline ordered and disordered phases. In this study, we investigate the different quantum phases that arise in ladder arrays comprising up to 92 neutral-atom qubits and experimentally observe the emergence of the quantum floating phase. We analyze the site-resolved Rydberg state densities and the distribution of state occurrences. The site-resolved measurement reveals the formation of domain walls within the commensurate ordered phase, 
which subsequently proliferate and give rise to the floating phase with incommensurate quasi-long-range order.
By analyzing the Fourier spectra of the Rydberg density-density correlations, we observe clear signatures of the incommensurate wave order of the floating phase. Furthermore, as the experimental system sizes increase, we show that the wave vectors approach a continuum of values incommensurate with the lattice. Our work motivates future studies to further explore the nature of commensurate-incommensurate phase transitions and their non-equilibrium physics.

\end{abstract}

\maketitle
The study of quantum phases and quantum phase transitions is one of the central topics in condensed matter, atomic, and high energy physics. Among these, commensurate to incommensurate phase transitions have been actively investigated since the 1980s~\cite{ HuseFisherPRL1982, PhysRevB.24.398, howes1983quantum, haldane1983phase, huse1984commensurate, DroseDepinning2003} and have attracted renewed attention recently~\cite{PhysRevLett.122.017205,rader2019floating,Chepiga&Mila2021Kibble,PhysRevResearch.4.043102, eck2023critical, PhysRevLett.124.103601}. 
A pivotal question revolves around the nature of these transitions and whether there exists an intermediate incommensurate phase when a commensurate solid melts into the disordered phase. In the context of spontaneous symmetry breaking, commensurate crystalline states in lattice systems often give hints to the underlying universality class of the phase transitions, such as the Ising and Potts classes \cite{WuPottsRev.1982}. 
Nonetheless, Huse and Fisher~\cite{HuseFisherPRL1982} proposed that when different kinds of domain walls exist in commensurate phases, they can introduce chiral perturbations, where the sequence of the domains matters. In such cases, several other possibilities could arise for the phase transition between the disordered and the commensurate ordered phases. These include a direct chiral transition \cite{HuseFisherPRL1982}, a direct first-order transition \cite{huse1984commensurate},
or an intriguing two-step transition across an incommensurate density-wave-order phase, i.e., the quantum floating phase~\cite{huse1984commensurate,PBak_1982}.

Early efforts in this field focused on probing classical floating phases in the melting of two-dimensional solids on periodic substrates~\cite{PhysRevB.49.2691, PhysRevB.49.2706, SchusterPT1996}. Later on, Fendley et al.~\cite{PhysRevB.69.075106} proposed a one-dimensional (1D) constrained hard-boson model that hosts a critical quantum floating phase. The constrained model is of particular relevance to the Rydberg atom array quantum simulation platform, which has made significant advancements in recent years~\cite{Bernien2017Dynamics, Keesling2019Kibble, Labuhn2016RydIsing,Léséleuc2019topo,Ebadi2021_256,Pascal2021AF}. Strong blockade interactions between nearby Rydberg atoms directly mimic the constraint in the hard-boson model. Theoretical calculations predict that critical quantum floating phases should exist in the 1D Rydberg atom array system \cite{PhysRevLett.122.017205,rader2019floating,Chepiga&Mila2021Kibble,PhysRevResearch.4.043102, Weimer2010twostage}. However, probing the floating phase in the 1D Rydberg system has been proven to be challenging due to their existence in only narrow regions in the phase diagram.

\begin{figure*}[t]
\centering
\includegraphics[width=1\textwidth]{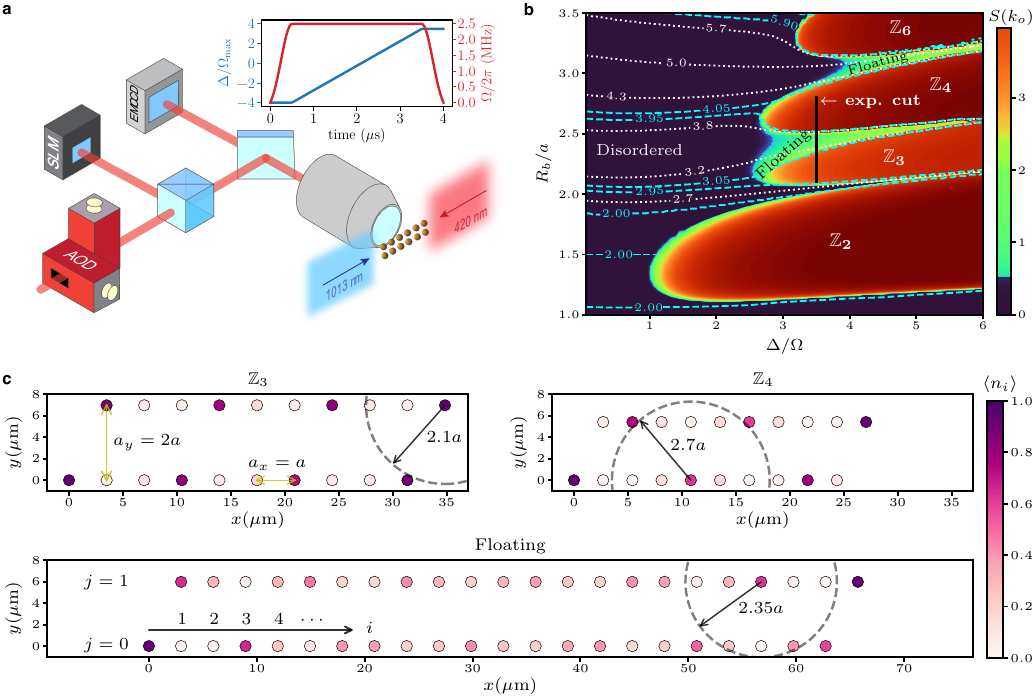}
\caption{\textbf{Quantum phases of Rydberg atoms arranged in a two-leg ladder.} \textbf{a}, Atoms are loaded into a two-leg ladder of optical tweezer traps generated using a SLM and rearranged into defect-free patterns by a second set of moving tweezers using a pair of crossed AODs. Coherent transitions are driven between the ground state $\ket{g} = \ket{5S_{1/2}}$ and the Rydberg state $\ket{r} = \ket{70S_{1/2}}$ in each atom with a two-photon transition induced by lasers at $420$ nm and $1013$ nm. The inset shows a linear detuning sweep $\Delta (t)$ at a constant Rabi frequency $\Omega_{\text{max}} = 2 \pi \times 2.5 \, \text{MHz}$ for preparing the ground states of the phase diagram via adiabatic evolution. Projection of the many-body quantum state into bitstrings of $\ket{g}$ and $\ket{r}$ for each atom can be detected on an EMCCD camera with the Rydberg state $\ket{r}$ detected as loss of atom. \textbf{b}, The ground-state phase diagram for the Rydberg Hamiltonian [Eq.~\eqref{eq:2legrydham}] in a two-leg ladder is shown with lattice spacings $a_x = a$ and $a_y = 2a$. Structure factors $S(k)$ are numerically computed for $1 \le R_b/a \le 3.5$ using DMRG (Supplementary Information). The color map depicts the peak height $S(k_o)$ at $k_o=2\pi/p$ with $p$ being the wavelength in units of the lattice constant $a$, while contour lines show the constant-$p$ lines. The $\mathbb{Z}_p$ orders have constant values of integer $p$, while the floating phase exhibits a continuously varying $p$. The black line cut corresponds to experimental parameters chosen in subsequent figures. \textbf{c}, Experimentally measured Rydberg densities illustrate the $\mathbb{Z}_p$ orders and the floating phase. The radius of the dashed circle illustrates the Rydberg blockade radius $R_b$. The $\mathbb{Z}_3$ and $\mathbb{Z}_4$ orders exhibit Rydberg density oscillations with periods of $p=3$ and $p=4$ lattice spacings, respectively. The incommensurate floating phase displays no discernible periodicity in density oscillations.}
\label{fig:phasediag}
\end{figure*}

In this study, we construct a quasi-1D array of \ce{^{87}Rb} atoms arranged in a two-leg ladder and experimentally probe the quantum floating phase. The Rydberg interactions between the two legs of the ladder array in the chosen geometric aspect ratio introduce stronger chiral perturbations (Supplementary Information) that yield broad regions of the floating phase in experimentally accessible parameter regimes (Fig.~\ref{fig:phasediag}\textbf{b}), which greatly facilitate our experimental observations. We provide supporting numerical calculations for the full phase diagram of the ladder system and experimentally measure the various phase regimes. By taking snapshots of the prepared states, we obtain the site-resolved Rydberg density and correlation functions. The dominant wave vectors for the measured correlation functions, extracted through Fourier analysis, offer a clear distinction between commensurate and incommensurate phases. An important feature of the floating phase is that, in the thermodynamic limit, the incommensurate wave vectors continuously depend on the physical parameters~\cite{PBak_1982}. We experimentally observe that, as system sizes increase, the incommensurate wave vectors tend towards a continuum of values.

\begin{figure*}[t]
\centering
\includegraphics[width=1\textwidth]{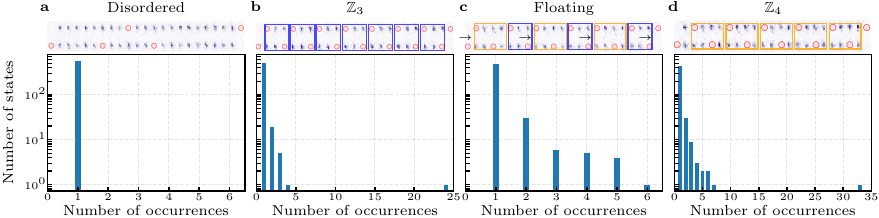}
    \caption{\textbf{Histogram of the measured bitstring occurrence frequency.} The experimental parameters for $\mathbb{Z}_3$, floating, and $\mathbb{Z}_4$ phases are along the black line shown in Fig.~\ref{fig:phasediag}\textbf{b} with $\Delta/\Omega = 3.5$. The atoms are arranged in a $(2 \times 21 +2)$ ladder array for all phases with $L = 21$. Top plots are fluorescence images of the $\ket{g}$ and $\ket{r}$ bitstrings with the largest occurrence for each phase, where atoms excited to \ket{r} are detected as atom loss and marked with red circles. \textbf{a}, The disordered phase at $\Delta/\Omega = 0$, $R_b/a = 2.4$ is sampled $573$ times, with each measured bitstring occurring only once. \textbf{b}, The $\mathbb{Z}_3$ order at $R_b/a = 2.1$ exhibits a perfect $\mathbb{Z}_3$ state occurring 24 times out of $588$ samples, significantly more frequent than all other states. Blue boxes encircle the $\mathbb{Z}_3$ unit cells. \textbf{c}, The floating phase at $R_b/a = 2.35$ is measured $600$ times, with $\mathbb{Z}_4$ (orange boxes) and $\mathbb{Z}_3$ (blue boxes) cells appearing alternatively in the most frequent state. Right arrows point to the domain walls where the $\mathbb{Z}_4$ domain changes sublattices (the index $i$ modulo $4$ changes for the rungs containing Rydberg states). \textbf{d}, The $\mathbb{Z}_4$ phase at $R_b/a = 2.7$ is sampled $599$ times, with perfect $\mathbb{Z}_4$ states occurring most frequently at $33$ times. Orange boxes encircle the $\mathbb{Z}_4$ unit cells. }
    \label{fig:histogram}
\end{figure*}

In order to prepare this geometry of atom arrangement, \ce{^{87}Rb} atoms are loaded from a magneto-optical trap into a two-dimensional (2D) array of optical tweezers generated using a spatial light modulator (SLM). We then rearrange the initially loaded atoms into a defect-free two-legged pattern using a second set of optical tweezers generated by a pair of crossed acousto-optical deflectors (AODs)~\cite{wurtz2023aquila}. In our system, qubits are encoded in the electronic ground state $\ket{g}=\ket{5S_{1/2}}$ and the Rydberg state $\ket{r} = \ket{70S_{1/2}}$. The transition between the two states is driven by a two-photon process with two counter-propagating laser beams at 420 nm and 1013 nm shaped into light sheets (Fig.~\ref{fig:phasediag}\textbf{a}). The coupling between the states and van der Waals interaction between Rydberg states result in the Hamiltonian
\begin{eqnarray}
\label{eq:2legrydham}
\nonumber \frac{\hat{H}}{\hbar} &=& \sum_{i=1}^{L} \sum_{j =0}^{1} \left( \frac{\Omega}{2}  \ket{g_{i,j}}\bra{r_{i,j}} + \text{h.c.} -\Delta \hat{n}_{i,j} \right)  \\
&+&  \sum_{\mathbf{r} \neq {\mathbf{r'}}} V_{\mathbf{r}, \mathbf{r'}} \hat{n}_{\mathbf{r}}\hat{n}_{\mathbf{r'}},
\end{eqnarray}
where $i = 1, 2, ..., L$ and $j = 0, 1$ are the rung index and the leg index, respectively, $\Omega$ is the effective two-photon Rabi frequency, $\Delta$ is the two-photon detuning, $\hat{n}_{i,j} = \ket{r_{i,j}}\bra{r_{i,j}}$ is the Rydberg density operator, and $V_{\mathbf{r}, \mathbf{r'}} = C_6/\lvert \mathbf{r} - \mathbf{r'} \rvert^{6}$ is the van der Waals interaction where $C_6 =  2\pi \times 862690 $ MHz $\mu \text{m}^6$ and $\mathbf{r} = i a_{x}\mathbf{e_x} + j a_{y}\mathbf{e_y}$.  In this work, we set the lattice spacings $a_x = a$ and $a_y = 2a$ (Fig.~\ref{fig:phasediag}\textbf{c}). The interactions are parameterized by the Rydberg blockade radius $R_b = (C_6/\Omega)^{1/6}$, within which the interaction is much larger than the Rabi frequency and it is thus energetically unfavorable to have more than one Rydberg excitation.

Complex many-body ground states emerge from the interplay between the detuning and the Rydberg interactions (Fig.~\ref{fig:phasediag}$\textbf{b}$). For small positive values of detuning $\Delta/\Omega$, the system has a low Rydberg state occupancy and displays a featureless disordered ground state. When the detuning becomes large, the Rydberg excitations can occupy one of every $p$ consecutive sites, with $p$ being an integer number and its value determined by $R_b/a$. Such ordered $\mathbb{Z}_p$ density wave states have been observed experimentally both in 1D and 2D Rydberg arrays \cite{Bernien2017Dynamics, Keesling2019Kibble,Ebadi2021_256}. In the intermediate values of $R_b/a$ between two different crystalline orders, the proliferation of different types of domain-wall excitations destroys the crystalline orders, but it can still stabilize a Rydberg density wave order with a wavelength that is incommensurate with the lattice spacing $a$ \cite{HuseFisherPRL1982,Chepiga&Mila2021Kibble}, giving rise to the quantum floating phase. Those density wave states have degeneracies related by translation and mirror symmetries. To facilitate experimental observations, we apply special boundary conditions to add two extra sites on the edges of the ladder array as shown in Fig.~\ref{fig:phasediag}$\textbf{c}$, which break the aforementioned symmetries (Supplementary Information). 

\begin{figure*}[t]
\centering
\includegraphics[width=1\textwidth]{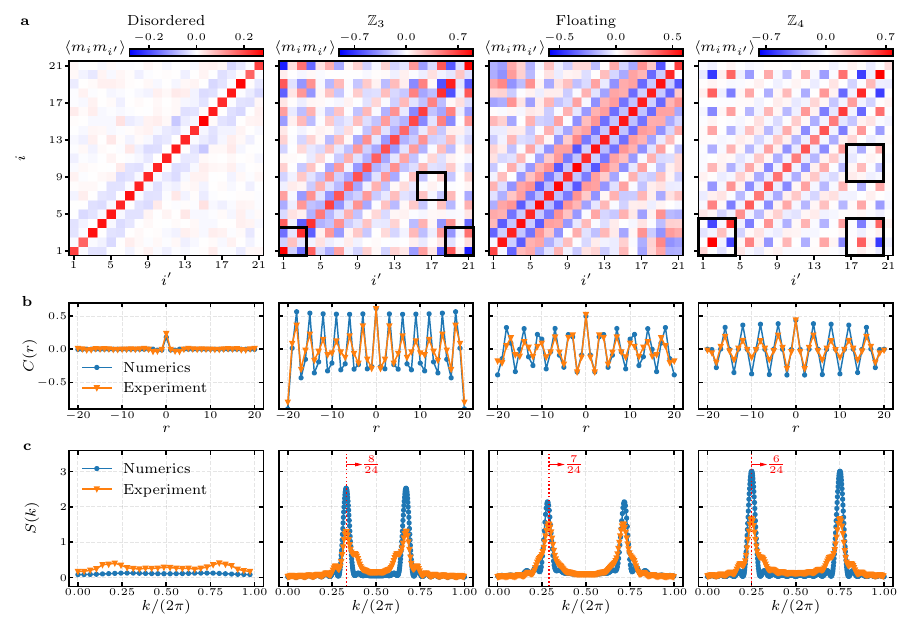}
\caption{\textbf{Correlation functions and structure factors for different phases.} \textbf{a}, The experimental correlation matrix $\langle m_i m_{i'} \rangle$ employing the order parameter $m_i = n_{i,1}-n_{i,0}$ for a system size of $L=21$. The presented results correspond to $(\Delta/\Omega, R_b/a) = (0, 2.4)$ in the disordered phase, $(3.5, 2.1)$ in the $\mathbb{Z}_3$ phase, $(3.5, 2.45)$ in the floating phase, and $(3.5, 2.7)$ in the $\mathbb{Z}_4$ phase. Black boxes indicate examples of minimal repeating patterns for $\mathbb{Z}_3$ and $\mathbb{Z}_4$ orders, which repeat every three or four sites, respectively. The floating phase lacks a repeating pattern due to an incommensurate wavelength. \textbf{b}, The mean correlator $C(r)$ is determined by averaging correlation functions $\langle m_i m_{i'} \rangle$ over the same relative distance $r=i-i'$. Both numerical and experimental results are presented, showing nearly identical oscillation periods. \textbf{c}, The structure factor $S(k)$ is derived from the Fourier transform of $\langle m_i m_{i'} \rangle$ with respect to $i-i'$. While weak signals are observed in the disordered phase, both numerical and experimental results display robust matching signals for $\mathbb{Z}_3$, floating, and $\mathbb{Z}_4$ phases.}
\label{fig:corrsl21}
\end{figure*}

The Rydberg density-density correlation functions can reflect (quasi) long-range orders in density-wave states and the Fourier analysis can extract the dominant wave vectors of these orders. To quantitatively map out the phase diagram, we use the following structure factor:
\begin{eqnarray}
\label{eq:strucfactor}
S(k) = \frac{p^2}{L^2} \sum_{i,i'} e^{\mathrm{i} k \left(i-i'\right)} \langle m_i m_{i'}\rangle,
\end{eqnarray}
where $m_i = n_{i,1}-n_{i,0}$ is the difference between the Rydberg density on the two sites in the $i$th rung and $p^2/L^2$ is an overall normalization factor. Here, $p$ is determined by $p = 2\pi / k_o$, where $L^2S(k)$ peaks at $k_o$. The ordered $\mathbb{Z}_p$ phase and the incommensurate floating phase correspond to $p$ being integer and non-integer values, respectively, and $S(k_o)$ converges to non-zero constant values as the system sizes increase. On the other hand, $S(k_o)$ goes to zero in the thermodynamic limit for the disordered phase.
The phase diagram for our ladder system is shown in Fig.~\ref{fig:phasediag}\textbf{b} by numerical calculations via density matrix renormalization group (DMRG). The maximum value $S(k_o)$ is plotted as a function of the physical parameters $\Delta/\Omega$ and $R_b/a$. The colormap shows clear differences among commensurate $\mathbb{Z}_p$ ordered phases, incommensurate floating phases, and the disordered phase. Lines of constant $p$ are superimposed to the phase diagram in Fig.~\ref{fig:phasediag}\textbf{b}. They reveal the wavelength of the density-wave order and thus whether it is commensurate or incommensurate. As illustrated, we can see the clear differences between commensurate $\mathbb{Z}_p$ ordered phases (with $p=2, 3, 4, 6$) and the incommensurate floating phase that lies between $\mathbb{Z}_3$ and $\mathbb{Z}_4$ orders or between $\mathbb{Z}_4$ and $\mathbb{Z}_6$ orders. The $\mathbb{Z}_5$ order and in fact $\mathbb{Z}_p$ orders for any odd $p \geq 5$ do not exist due to strong inter-leg Rydberg blockade interactions (Supplementary Information). We emphasize that determining the phase diagram by mapping out the structure factors, as demonstrated here, has the desirable benefit of being directly measurable in our experiments, in contrast to using the entanglement entropy (Supplementary Information).

Our experimental setup allows many repetitions of the adiabatic preparation process, giving us direct access to snapshots of bitstrings of each atom in $\ket{g}$ and $\ket{r}$ in each observation. We plot the experimentally measured density profiles for $\mathbb{Z}_3$, $\mathbb{Z}_4$, and floating phases in Fig.~\ref{fig:phasediag}\textbf{c}. In each experimental run, the corresponding many-body ground state is prepared by an adiabatic evolution starting from all atoms in the $\ket{g}$ state, followed by an adiabatic sweep similar to the one shown in the inset of Fig.~\ref{fig:phasediag}\textbf{a}. One can see that the $\mathbb{Z}_p$ order has large occupation of Rydberg states every $p$-th site on each leg, while the Rydberg density in the floating phase does not have the same repetitive structure in the array. Analysis of the statistical distributions of the read-out configurations provides clear indications of the different many-body phases prepared (Fig.~\ref{fig:histogram}). In the case of the disordered state with  $\Delta/\Omega = 0$, the measured bitstring configurations are likely to be all different, i.e., all the bitstrings occur only once (see Fig.~\ref{fig:histogram}\textbf{a}, with a total of $573$ samples taken in our experiments). A typical measurement read-out shows no order along the ladder, where a representative snapshot is shown in the florescence image above the histogram plot. 
In contrast, for the commensurate $\mathbb{Z}_3$ ($588$ samples) and $\mathbb{Z}_4$ ($599$ samples) phases, there is only one particular configuration that occurs much more frequently than all the other configurations (Fig.~\ref{fig:histogram}\textbf{b} and \ref{fig:histogram}\textbf{d}), corresponding to the classical density-wave state with periodicities $p=3$ and $p=4$, respectively, as shown in the florescence images.

More interestingly, within the floating phase situated between the  $\mathbb{Z}_3$ and $\mathbb{Z}_4$ phase regimes, the most frequently observed bitstring out of $600$ samples does not exhibit $\mathbb{Z}_3$ or $\mathbb{Z}_4$ periodicity. Instead, there are always smaller regions with $\mathbb{Z}_3$ or $\mathbb{Z}_4$ order appearing alternately, as shown by the alternate orange and blue boxes in the florescence image in Fig.~\ref{fig:histogram}\textbf{c}. This observation reveals that the floating phase is not simply a macroscopic superposition of $\mathbb{Z}_3$ and $\mathbb{Z}_4$ orders. If the melting of the $\mathbb{Z}_4$ order is considered, the domain walls can be identified by observing where the $\mathbb{Z}_4$ domain changes sublattices (Supplementary Information). These domain walls, as indicated by the right arrows, are essential for the incommensurate nature of the quantum floating phase (Fig.~\ref{fig:phasediag}\textbf{c}). In principle, the domain walls should be distributed at equal distances across the entire system for the ground state. However, changing the position of the domain walls incurs only small energy cost in the floating phase, which results in a nearly continuous distribution of states with high occurrences~\footnote{In the thermodynamic limit, a continuum limit can be defined and continuous translations can be performed on each configuration without any cost in energy. This translation corresponds to an overall phase shift of the incommensurate density-wave order, and the degeneracy is a manifestation of an emergent global U(1) symmetry.}. We observe such a feature in the bitstring distribution in Fig.~\ref{fig:histogram}\textbf{c}. This behaviour stands in contrast to the commensurate ordered phases, where only one bitstring occurs dominantly. These experimental results are fully consistent with theoretical calculations (see the Supplementary Information for numerical results).

\begin{figure*}
\centering
\includegraphics[width=1\textwidth]{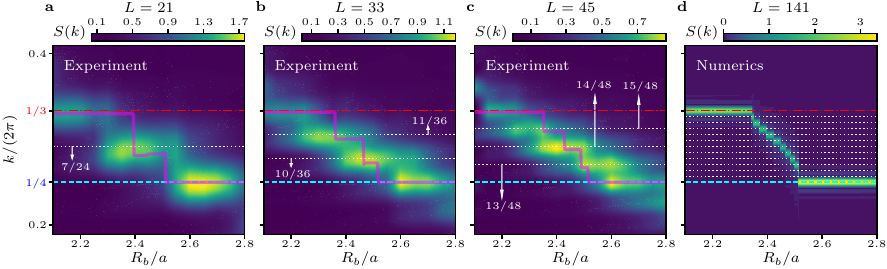}
\caption{\textbf{The structure factor $S(k)$ for different system sizes.} \textbf{a} - \textbf{c}, Experimental measurements of $S(k)$ are taken along the $\Delta/\Omega=3.5$ line, as labeled by the black line in Fig.~\ref{fig:phasediag}\textbf{b}, for system sizes $L=21, 33, 45$. The data points cover $R_b/a$ values ranging from $2.1$ to $2.8$ with a step size of $0.05$. The magenta curves represent the peak positions of $S(k)$ obtained from numerical calculations. The experimental data show that the number of peak plateaus allowed in the quantum floating phase increases with system sizes. \textbf{d}, Numerical results for $L=141$ demonstrate a continuous change in the wave vector of the quantum floating phase in the large $L$ limit.}
\label{fig:strucfactor}
\end{figure*}

The $\mathbb{Z}_3$, $\mathbb{Z}_4$, and intermediate quantum floating phases can be further elucidated through a quantitative analysis of the Rydberg density correlations. In Fig.~\ref{fig:corrsl21}$\textbf{a}$, we present the correlation matrix $\langle m_i m_{i'}\rangle$ between rungs $i$ and $i'$. In the disordered phase, the diagonal elements of the correlation matrix are non-zero, while the off-diagonal elements quickly decay to zero, accompanied by weak site-dependent fluctuations. This behavior indicates that the correlation length is short in this case. Quantitatively, by averaging the correlation functions over sites with a fixed relative distance $i-i'$, we show that the mean correlation function $C(r) = 1/N_r \sum_{i-i'=r} \langle m_i m_{i'}\rangle$ ($N_r$ is the number of $(i, i')$ pairs satisfying $i-i'=r$) decays to zero immediately for $|i-i'| > 0$ (first column of Fig.~\ref{fig:corrsl21}$\textbf{b}$). The numerical results are in good agreement with experimental measurements. Both experimental and numerical structure factors $S(k)$ show a weak signal for density fluctuations, as shown in the first column of Fig.~\ref{fig:corrsl21}$\textbf{c}$. 

The same analysis is performed for $\mathbb{Z}_3$, floating, and $\mathbb{Z}_4$ phases. The integer-period oscillation patterns for $\mathbb{Z}_3$ and $\mathbb{Z}_4$ phases can be easily read from the correlation matrix, where the correlations oscillate with periodicities of $p=3$ and $p=4$, respectively. On top of the periodic oscillations, there also exists some amplitude decay for the correlations as the distance increases. 
The black boxes in Fig.~\ref{fig:corrsl21}$\textbf{a}$ indicate the smallest repeating units: $3 \times 3$ and $4\times 4$ squares for $\mathbb{Z}_3$ and $\mathbb{Z}_4$ orders, respectively. There is no such repeating unit in the correlation matrix for the quantum floating phase, indicating that its periodicity is not commensurate with the lattice spacing. The mean correlation function in Fig.~\ref{fig:corrsl21}$\textbf{b}$ also shows perfect $\mathbb{Z}_3$ and $\mathbb{Z}_4$ oscillations for the commensurate phases, while it presents an incommensurate oscillation pattern for the floating phase. The plots of $C(r)$ exhibit excellent agreement with numerical results in the short range, albeit with amplitude decay due to the finite correlation length in our experiments. The experimental data for the three density-wave orders show strong signals of oscillations, which are absent in the disordered phase.

The structure factor in Fig.~\ref{fig:corrsl21}$\textbf{c}$ has peaks at $2\pi\times 8/24$, $2\pi\times 7/24$, and $2\pi\times 6/24$ for the $\mathbb{Z}_3$, the incommensurate floating, and the $\mathbb{Z}_4$ phases, respectively, in agreement with numerical results. These particular values for the peak positions can be explained by the finite size of the ladder and our choice of boundary conditions. With finite detuning, a Rydberg excitation is pinned at each of the two edge sites of the ladder since that minimizes the Rydberg blockade energy, as illustrated in Fig.~\ref{fig:phasediag}$\textbf{c}$. Under this boundary condition, the Rydberg density waves for $\mathbb{Z}_3$ and $\mathbb{Z}_4$ orders are both compatible with system sizes satisfying $L=12l-3$, with $l$ being any positive integer number. Because of the Rydberg interactions, the pinning of the Rydberg density wave at the edge sites on a finite lattice restricts arbitrary continuous variation of the wavelength in the floating phase, allowing only 
$l-1$ fractional values of wavelength between the $\mathbb{Z}_3$ and the $\mathbb{Z}_4$ orders. For example, in the $\mathbb{Z}_3$ phase ($p=24/8$) with a system size of $L=21$, the most frequently occurring state contains $8$ excitations in each leg, while in the $\mathbb{Z}_4$ phase ($p=24/6$), it comprises $6$ excitations in each leg. In the floating phase between the $\mathbb{Z}_3$ and $\mathbb{Z}_4$ phases, the most common state consists of $7$ excitations in each leg, resulting in a fractional $p=24/7$, as confirmed by the results in Fig.~\ref{fig:histogram}\textbf{c}. In the large $L$ limit, this series of $l-1$ waves with fractional wavelengths continuously interpolates between two crystalline orders with integer values $p=3$ and $p=4$. In the following, we experimentally explore this trend by constructing ladder arrays of different sizes.

We vary $R_b/a$ in small steps by changing the lattice constant $a$ and examine the changes in the structure factor $S(k)$ for various system sizes. This experimental cut is indicated by the black line in Fig.~\ref{fig:phasediag}\textbf{b} at $\Delta/\Omega=3.5$ and $R_b/a$ ranges from $2.1$ to $2.8$. $S(k)$ is measured for different system sizes $L=21, 33, 45$, with its dependence on $R_b/a$ and $k$ presented in Fig.~\ref{fig:strucfactor}$\textbf{a}$-$\textbf{c}$. We note that at each value of $R_b/a$, the corresponding many-body ground state is prepared by an adiabatic sweep starting from all atoms in the $\ket{g}$ state. As we can see, the structure factor exhibits $k/2\pi = 1/3$ and $1/4$ peak plateaus independent of system sizes, signifying the existence of $\mathbb{Z}_3$ and $\mathbb{Z}_4$ orders. More importantly, we experimentally observe, in the floating phase, $1$, $2$, and $3$ peak steps for $L=21$, $33$, and $45$, respectively, which are fully consistent with our theoretical expectations. Furthermore, the step locations are observed at the corresponding incommensurate values, also in agreement with theory results. In particular, we observe $k/2\pi=7/24$ for $L=21$, $k/2\pi=10/36, 11/36$ for $L=33$, and $k=13/48, 14/48, 15/48$ for $L=45$. Our experiments demonstrate that the number of incommensurate steps between crystalline orders increases with the system size, which agrees with the theoretical expectation of its tending towards infinity as the system approaches the thermodynamic limit. Consequently, the dominant wave vector of the quantum floating phase would change continuously with the physical parameters. An example of the structure factor in the large $L$ limit is shown in Fig.~\ref{fig:strucfactor}$\textbf{d}$ using DMRG calculations for $L=141$. In this case, the dominant wave vector in the floating phase takes a series of values, nearly forming a continuous curve within a small window between $R_b/a=2.34$ and $R_b/a=2.51$. In addition, we conducted measurements in the same $R_b/a$ range with fixed $\Delta/\Omega = 0$, where the system is always in the disordered phase. The results show very weak peaks in the structure factor, and the signals diminish with increasing system sizes (Supplementary Information). These vanishing signals for the disordered phase confirm that the boundary effects do not induce artificial signatures for the existence of density-wave orders in our experiments, and, in particular, the observed peaks in the incommensurate wave vectors are genuine features of the quantum floating phase.

In summary, our utilization of Rydberg-atom ladder arrays has enabled the experimental observation of an incommensurate density-wave order—the quantum floating phase. Distinct Rydberg density fluctuations are observed in different density-wave states, and the structure factor exhibits strong signatures for the existence of the floating phase. Through systematic tuning of the lattice constant, discrete wave vectors allowed in the quantum floating phase for different system sizes are observed, indicating their convergence to a continuum of incommensurate values in the thermodynamic limit. This experiment underscores the versatility of Rydberg atoms in ladder arrays as highly programmable quantum simulators, facilitating the exploration across various quantum many-body phases. The constructed Rydberg arrays house a broad spectrum of critical lines and points in experimentally accessible regimes, including Ising, Potts, Ashkin-Teller, chiral, Berezinskii–Kosterlitz–Thouless \cite{Kosterlitz_1974}, Pokrovsky-Talapov \cite{Pokrovsky1979}, and Lifshitz critical points \cite{PhysRevResearch.3.023049} (see Supplementary Information for details). The $\mathbb{Z}_4$ regime in our system establishes an ideal platform for investigating the Kibble-Zurek mechanism of chiral phase transitions. Our work paves the way for future experimental inquiries into these critical phenomena, and it offers potential applications in diverse contexts, such as quantum simulation of lattice gauge theories \cite{Bazavov:2015kka,Zhang:2018ufj,Surace:2019dtp,meurice2021theoretical, Heitritter:2022jik}, inhomogeneous phases, the Lifshitz regime of lattice quantum chromodynamics \cite{Pisarski:2019cvo, Kojo:2009ha}, and ``chiral spiral” condensation in interacting fermionic systems \cite{Basar:2008im}.
\begin{acknowledgments}

We thank K. Heitritter, J. Corona, Milan Kornjača, and members of the QuLAT collaboration for helpful discussions, and Simon Evered for providing helpful feedback on the manuscript. Authors from QuEra Computing Inc.~acknowledge the support from the DARPA ONISQ program (grant no.~W911NF2010021).
S.-W.T. acknowledges the support from the National Science Foundation (NSF) RAISE-TAQS under Award Number 1839153. J.Z.\ is supported by
NSFC under Grants No.~12304172 and No.~12347101, Chongqing Natural Science Foundation under Grant No.~CSTB2023NSCQ-MSX0048, and Fundamental Research Funds for the Central Universities under Projects No.~2023CDJXY-048 and No. 2020CDJQY-Z003. Y.M.~and, in the early stage of the project, J.Z. were supported in part by the U.S.~Department of Energy (DoE) under Award Number DE-SC0019139.
Computations were performed using the computer clusters and data storage resources of the UCR High Performance Computing Center (HPCC), which were funded by grants from NSF (MRI-1429826) and NIH (1S10OD016290-01A1). 

\end{acknowledgments}

%

\vfill\eject
\clearpage

\onecolumngrid

\begin{center}

{\large \bf Supplementary Information to:\\
Probing quantum floating phases in Rydberg atom arrays}\\

\vspace{0.6cm}

Jin Zhang$^{1,2}$, Sergio H. Cant\'u$^{3}$, Fangli Liu$^{3}$, Alexei Bylinskii$^{3}$, Boris Braverman$^{3}$, Florian Huber$^{3}$, Jesse Amato-Grill$^{3}$, Alexander Lukin$^{3}$, Nathan Gemelke$^{3}$, Alexander Keesling$^{3}$, Sheng-Tao Wang$^{3}$, Y. Meurice$^1$, S.-W. Tsai$^4$ \\

$^1${\it Department of Physics and Astronomy, The University of Iowa, Iowa City, IA 52242, USA}

$^2${\it Department of Physics and Chongqing Key Laboratory for Strongly Coupled Physics, Chongqing University, Chongqing 401331, China}

$^3${\it QuEra Computing Inc., 1284 Soldiers Field Road, Boston, MA, 02135, USA}

$^4${\it Department of Physics and Astronomy, University of California, Riverside, CA 92521, USA}
 \end{center}

\vspace{0.6cm}
\tableofcontents

\newpage 

\twocolumngrid

\section{Methods}

\subsection{Creation of Atom Array}
In this experiment, we utilize \ce{^{87}Rb} atoms captured from a magneto optical trap (MOT). The atoms are loaded with $59\%$ probability into optical tweezers generated using 840 nm light, which are created through a phase-only spatial light modulator (SLM, Hamamatsu X15213 series) to generate the trapping potentials with a depth of 10 MHz, as determined by the Stark shift associated with the $\ket{5S_{1/2}, F=2, m_{F}=2}$ to $\ket{5P_{3/2}, F=3, m_{F}=3}$ transition. Subsequently, we image atoms for 15 ms using near resonant 780 nm light, which is captured through a microscope objective of $\text{NA} =0.65$ to image onto an EMCCD camera (Andor iXon Ultra) to capture the occupation of each site with $99.5\%$ fidelity. From the atom detection results, atoms are moved using acousto-optic deflectors (AODs) to sort atoms onto the target two-legged pattern with $99.5\%$ sorting fidelity. Following imaging, the atoms are cooled to a temperature of 10 $\mu$K and polarized into the $\ket{5S_{1/2}, F=2, m_{F}=2}$ state. 

\subsection{Rydberg System}

The atomic levels we use in our qubit system are $\ket{g} = \ket{5S_{1/2},F=2,m_F=-2}$ and $\ket{r} = \ket{70S_{1/2},m_j=-1/2,m_I=-3/2}$, which are coherently driven by a two-photon transition induced by lasers at $420$ nm and $1013$ nm. The 420 nm laser is a frequency-doubled Ti:Sapphire laser (M Squared). We frequency stabilize the laser by locking a frequency sideband generated by an electro-optic modulator to an ultra-low-expansion (ULE) reference cavity (notched cylinder design, Stable Laser Systems), with finesse $\mathcal{F} = 30,000$ at 840 nm. The 1,013 nm laser source is a whispering-gallery mode laser (OEwaves - OE3745), which is locked to the same reference cavity with a finesse of $\mathcal{F}=50,000$. The 1013-nm laser is amplified by an fiber amplifier (IPG YAM-100-1013-LP-SF).

We place the Rydberg two-photon transition at a single-photon detuning of $\delta \approx 2\pi \times 1$ GHz blue detuned from the intermediate state $\ket{6P_{3/2}}$ with Rabi frequencies $\left(\Omega_{420},\Omega_{1013}\right) = 2\pi \times (114,44)$ MHz, and a two-photon Rabi frequency of $\Omega = \Omega_{420}\Omega_{1013}/2\delta \approx 2\pi \times 2.5$ MHz. 

\subsection{Rydberg Pulses}
Atoms are adiabatically released from the tweezer traps after initializing them into the ground state $\ket{g}$. The total time with the traps off is 7 $\mu$s. It is during the trap-off time that we apply a Rydberg pulse. The pulse is described by a time-dependent Rabi frequency $\Omega\left(t\right)$, time-dependent detuning $\Delta\left(t\right)$, and instantaneous phase $\phi\left(t\right)$. This is implemented by controlling the amplitude, frequency, and phase of the 420-nm laser using a double-pass acousto-optic modulator (AOM) configuration.

\subsection{Rydberg Beam Shape}
We shape both Rydberg excitation beams into one-dimensional top-hats (light sheets) to homogeneously illuminate the plane of atoms. This is done by placing a SLM in the Fourier plane of each Rydberg beam. This allows us to control both phase and intensity on the atoms at the cost of efficiency. The tophat geometry is 75 $\mu$m wide in the plane of the atom array, and has a waist of $35\,\mu$m along the axis normal to the atom array plane.

\begin{figure*}[t]
\centering
\includegraphics[width=1\textwidth]{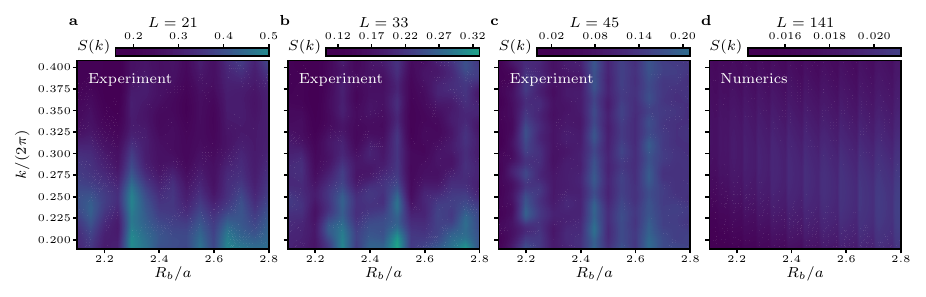}
\caption{\textbf{The structure factor $S(k)$ in the disordered phase for different system sizes.} \textbf{a} - \textbf{c}, Experimental measurements of $S(k)$ are conducted along the $\Delta/\Omega=0$ cut, corresponding to the disordered phase. The peaks of $S(k)$ are significantly smaller compared to those in Figs.~\ref{fig:strucfactor}\textbf{a - c} for the same $L$, and no signals of density wave order are observed. The numerical results for $L=141$ indicate that $S(k)$ tends to zero in the disordered phase in the large $L$ limit. }
    \label{fig:expdisordered}
\end{figure*}

\subsection{Experimental Results for the Disordered Phase}
We have studied the dependence of the structure factor $S(k)$ on the Rydberg blockade radius $R_b/a$ in Fig.~\ref{fig:strucfactor} in the main text, which shows strong signatures for the existence of the crystalline $\mathbb{Z}_3$ and $\mathbb{Z}_4$ orders and the incommensurate floating phase. In order to preclude the possibility that the signals are introduced by the boundary effects, we also experimentally measure $S(k)$ in the disordered phase at $\Delta/\Omega = 0$ for the same range of $R_b/a$ between $2.1$ and $2.8$ and the same system sizes $L = 21, 33, 45$. The results are presented in Figs.~\ref{fig:expdisordered}\textbf{a}-\textbf{c} with a numerical result for $L=141$ in Fig.~\ref{fig:expdisordered}\textbf{d}. Very weak signals can be seen in $S(k)$ for all the cases in Fig.~\ref{fig:expdisordered}. More importantly, there is no dominant peak in $S(k)$, so no indication of the presence of a density wave order in the prepared states. These signals in $S(k)$ diminish as the system sizes increase, indicating that the structure factor for the disordered phase tends to be zero in the thermodynamic limit. The numerical result shows that the peak height of $S(k)$ for $L=141$ is about $1/25$ of that for $L=21$, which is almost zero and confirms that $S(k)$ for the disordered phase approaches zero in the large $L$ limit. Therefore, we conclude that the boundary effects will not induce artificial signatures for the existence of density-wave orders in our experiments.

\subsection{Numerical Methods}
We perform finite-size density-matrix renormalization group (DMRG) calculations \cite{PhysRevLett.69.2863, PhysRevB.48.10345,PhysRevLett.75.3537} with the \textsc{ITensor Julia Library} \cite{10.21468/SciPostPhysCodeb.4}. All the Rydberg interactions within $21$ consecutive rungs are kept in the Hamiltonian to maximally mimic the long-range interaction in the Rydberg Hamiltonian. When searching for the ground state, we gradually increase the maximal bond dimension $D$ during the variational sweeps until the truncation error $\epsilon$ is below $10^{-10}$. Some calculations using different truncation errors are specified later. DMRG sweeps are terminated once the ground-state energy changes less than $10^{-11}$ and the von Neumann entanglement entropy changes less than $10^{-8}$ in the last two sweeps. Generally, thousands of DMRG sweeps are needed in the quantum floating phase and more sweeps are needed for convergence for larger Rydberg blockade radius $R_b/a$.

\begin{figure*}[t]
\centering
\includegraphics[width=1\textwidth]{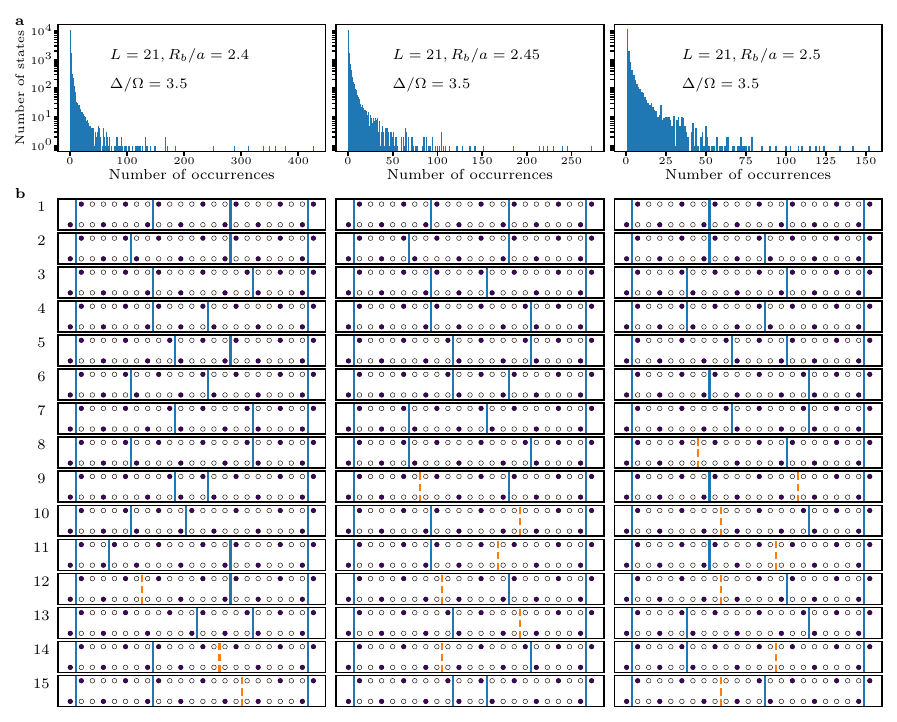}
\caption{\textbf{Bitstring occurrence frequency with $40,000$ numerical samples from the ground-state wavefunction.} \textbf{a}, Histograms of the number of occurrences for the sampled states are plotted in the quantum floating phase of the Rydberg ladder with $L=21, \Delta/\Omega=3.5$, and $R_b/a=2.4, 2.45, 2.5$. \textbf{b}, The most probable bitstrings are shown in descending order. Blue lines mark the domain walls with two Rydberg states in nearest-neighbor rungs but different legs. Orange dashed lines mark the domain walls with no Rydberg states in nearest-neighbor rungs.}
    \label{fig:numsampling}
\end{figure*}

\subsection{Simulation of Finite Experimental Measurement Samples}
In Fig.~\ref{fig:histogram} of the main text, our experimental data show that the distribution of bitstring occurrence frequency is very distinct in different phases and reveal that the quantum state of the floating phase is created by the proliferation of domain walls. Here, we provide numerical data to the experimental results for the floating phase. Figure~\ref{fig:numsampling} shows the numerical bitstring occurrence frequency for $L = 21$ and $\Delta/\Omega=3.5$ sampled from the many-body wavefunction. The values of $R_b/a = 2.4, 2.45$, and $2.5$ are inside the floating phase close to $\mathbb{Z}_3$ boundary, at the midpoint of $\mathbb{Z}_3$ and $\mathbb{Z}_4$ boundaries, and close to $\mathbb{Z}_4$ boundary, respectively. Notice that the floating phase in the experiment has slight shift relative to the numerical results and the experimental results for $R_b/a=2.35$ are presented in the main text. We calculate the ground-state wavefunction using DMRG, and sample the classical bitstrings $40,000$ times according to the density matrix. It is seen that the number of occurrences for the most probable state are 426, 272, and 152, respectively. Unlike the $\mathbb{Z}_p$ ordered state where the occurrence frequency of the classical $\mathbb{Z}_p$ state is much larger than other states (see Fig.~\ref{fig:histogram} in the main text), there are many classical bitstrings with similar occurrence frequencies that decrease slowly to zero. 

As explained in the main text and here below, the wave vector of the finite-size floating phase for $L=21$ takes one value $k=2\pi \times 7/24$. It is thus expected that the most probable classical bitstrings have 7 Rydberg excitations in each leg with domain walls almost evenly distributed in the ladder. The $15$ most probable bitstrings are shown in Fig.~\ref{fig:numsampling}\textbf{b}. One can see that the most probable bitstring is exactly the same as the one in experimental measurements. For numerical results, the seven most probable bitstrings for three values of $R_b/a$ are exactly the same, indicating that the three cases are in the same floating phase as expected. The six most probable bitstrings for the numerical results can also be found in the fifty most probable bitstrings for the experimental measurements. In each leg of these states, the seven Rydberg states arrange themselves such that ``rggg" and ``rgg" appear alternately. The orders of arrangements of the Rydberg states in the two legs are opposite to further reduce the energy. Notice that the classical $\mathbb{Z}_4$ state has only one Rydberg state in any two nearest-neighbor (NN) rungs (see Fig.~\ref{fig:histogram}\textbf{d}). We identify NN rungs with two Rydberg states on the diagonal sites as a domain wall (type-1, highlighted by blue solid lines in Fig.~\ref{fig:numsampling}\textbf{b}). This type of domain walls also labels where the $\mathbb{Z}_4$ domain changes sublattices. For instance, the first domain of the most probable state has Rydberg states on the rungs with indices $i_r \equiv 1,3 \pmod{4}$, while $i_r \equiv 0,2 \pmod{4}$ for the second domain and $i_r \equiv 3,1 \pmod{4}$ for the third domain. Thus, there are four domain walls equally distributed in the most probable bitstring that has the minimal energy. Nonuniform distributions of domain walls cost energy and have lower occurrence frequency. We have checked that twenty out of twenty-two most probable bitstrings for the experimental results have seven Rydberg excitations in each leg and four type-1 domain walls. So our experiment successfully probe the quantum floating phase up to some small defects.

Another type of domain wall shown in Fig.~\ref{fig:numsampling}\textbf{b} is the empty NN rungs (type-2, highlighted by orange dashed lines), which not only label where the $\mathbb{Z}_4$ domain changes sublattices but also reduce the number of Rydberg states and cost more energy. Thus, it appears in the states with relatively lower occurrence frequency. Notice that increasing $R_b/a$ brings more repulsive energy that tends to decrease the Rydberg density, so the occurrence frequency of the most probable bitstring that only have type-1 domain walls decreases with increasing $R_b/a$. In the large $L$ limit, the quantum state in the floating phase is a superposition of an infinite number of these domain-wall states. An infinitesimal redistribution of these domain-wall states gives gapless excitations.

\begin{figure}[t]
\centering
\includegraphics[width=\linewidth]{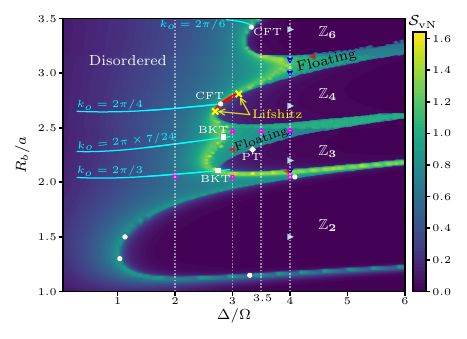}
\caption{\textbf{Ground-state phase diagram using von Neumann entanglement entropy.} The results  for $1 \le R_b/a \le 2.5$, $2.5 < R_b/a \le 3.15$, and $3.15 < R_b/a \le 3.5$ are computed for systems with $L = 288$, $285$, and $290$, respectively (Fig.~\ref{fig:phasediag}\textbf{b} in the main text also employs this technique). Slight variations in $L$ in different regimes ensure compatibility with the periods of the crystalline orders. The dark lobes represent crystalline orders, and the green areas between crystalline orders, bounded by the bright yellow lines, show the floating phase. The bright yellow lines are BKT transition lines, separating the floating phases and the disordered phase. The boundaries between the floating phases and the crystalline orders are PT transition lines. The red line labels the direct phase transition between the $\mathbb{Z}_4$ order and the disordered phase, which is a chiral transition line with continuously varying critical exponents plus a single CFT point (white circle). The chiral transition line terminates at two Lifshitz points (yellow cross) where the floating phases emerge. There also exist direct phase transitions between the $\mathbb{Z}_6$ order and the disordered phase, which also include a single CFT point. On the equal-$k_o$ lines (cyan lines) in the disordered phase, the peak position of the structure factor $S(k)$ remains constant at $k_o$. Commensurate lines, where $2\pi/k_o$ is an integer, intersect with the $\mathbb{Z}_{4(6)}$ boundary at the CFT point. The floating phase fully encompasses the $\mathbb{Z}_3$ order, and the $k_o=2\pi/3$ line goes from the disordered phase into a critical phase and then into the $\mathbb{Z}_3$ order (see Fig.~\ref{fig:bktpttransitions}). White circles on the $\mathbb{Z}_2$ boundary denote four Ising critical points studied in Fig.~\ref{fig:collapseu4gapising}. Triangles, squares, and the diamond represent the points discussed in the subsequent figures.}
    \label{fig:entphasediag}
\end{figure}

\section{Phase Diagram from Entanglement Entropy}

The phase diagram shown in Fig.~\ref{fig:phasediag}\textbf{b} is plotted from the structure factor $S(k)$, which is experimentally accessible by measuring the density-density correlations. Theoretically, the entanglement entropy is a universal tool to map out the phase diagram of quantum systems without any prior knowledge. Thus, we show numerical results below the corresponding phase diagram produced from entanglement entropy. We also discuss rich physics of quantum phase transitions and critical phenomena present in our Rydberg ladder system setup and provide substantial numerical evidence in the following sections.

For a quantum system consisting of $\mathcal{A}$ and $\mathcal{B}$ parts, the bipartite ground-state entanglement entropy is defined as $\mathcal{S}_{\rm{vN}}=-\Tr_{\mathcal{B}}\rho_{\mathcal{A}}\ln\rho_{\mathcal{A}}$, where the reduced density operator of subsystem $\mathcal{A}$ is $\rho_{\mathcal{A}}=\Tr_{\mathcal{A}}\left(\ket{\Psi_0}\bra{\Psi_0}\right)$ and $\ket{\Psi_0}$ is the ground state. At critical points of one-dimensional (1D) quantum systems with open boundary conditions (OBC) or periodic boundary conditions (PBC), the conformal field theory (CFT) predicts that the entanglement entropy has the following form \cite{AffleckCritical1991,HOLZHEY1994443,VidalEntangle2003,PasqualeCalabrese_2004,Calabrese_2009}
\begin{eqnarray}
\label{eq:ent}
\mathcal{S}_{\rm{vN}}= \begin{cases} \frac{c}{6} \ln \left\{\frac{4(L_s+1)}{\pi} \sin \left[\frac{\pi(2 l_a+1)}{2(L_s+1)}\right]\right\}+s_{o} & \textrm{for OBC} \\ \frac{c}{3} \ln \left[\frac{L_s}{\pi} \sin \left(\frac{\pi l_a}{L_s}\right)\right]+s_{p} & \textrm{for PBC}
\end{cases},
\end{eqnarray}
where $c$ is the central charge, $s_{o(p)}$ are constants, $L_s$ and $l_a$ are the sizes of the 1D system and the subsystem $\mathcal{A}$, respectively. For a fixed ratio $l_a/L_s$, the entanglement entropy between two parts diverges logarithmically with $L_s$, which can be used to pick out the critical lines (phases) in the phase diagram.

In our setup, the two-leg Rydberg ladder has lattice spacings $a_x = a$ and $a_y = 2a$ and it consists of $L$ rungs plus two extra sites on the bottom left corner and the top right corner, respectively. We refer this boundary condition in a ladder as the shifted boundary condition (SBC), while ladder systems with OBCs do not have the two extra sites. In DMRG calculations, we treat the ladder system as a ``chain with OBC'' and label the sites by integers from $1$ to $L_s=2L+2$. The two boundary sites are labeled by $1$ and $2L+2$, respectively. In the bulk, the sites in the bottom row are labeled by even numbers, while the sites in the top row are labeled by odd numbers. The entanglement entropy between two halves of the ladder is used to draw the phase diagram shown in Fig.~\ref{fig:entphasediag}. It is seen that the low entanglement areas precisely depict the crystalline orders with spontaneous $\mathbb{Z}_p$ ($p=2,3,4,6$) symmetry breaking. In 1D Rydberg chains, $\mathbb{Z}_p$ orders with integer $p \ge 2$ will successively show up as $R_b$ increases, while ladder systems may forbid certain orders due to strong interchain interactions. Here, $\mathbb{Z}_p$ orders with odd $p \ge 5$ are forbidden in our ladder system, as we explain in the next paragraph. Based on other studies for 1D Rydberg systems \cite{PhysRevB.69.075106,Weimer2010twostage,SelaDislocation2011,PhysRevLett.122.017205,rader2019floating}, the high $\mathcal{S}_{\rm{vN}}$ plateau between $\mathbb{Z}_3$ and $\mathbb{Z}_4$ orders and the one between $\mathbb{Z}_4$ and $\mathbb{Z}_6$ orders are the incommensurate density waves or the critical quantum floating phase with emergent U(1) symmetry studied in the main text. We can also identify the disordered phase in the regime that smoothly connects to the $\Delta/\Omega \rightarrow 0$ limit.

Strong interchain interactions significantly facilitate the formation of $\mathbb{Z}_p$ orders with even $p$. Given that the Rydberg state in one leg blockades the region with a length of $2\sqrt{R^2_b-4a^2}$ in the other leg, a $\mathbb{Z}_p$ order with even $p$ needs at most $R_b/a = \sqrt{p^2/4+4}$ to form, while the formation of $\mathbb{Z}_p$ orders in the 1D Rydberg chain requires $p-1 < R_b/a < p$. Consequently, crystalline orders in our ladder system manifest at a significantly smaller Rydberg blockade radius than those in the 1D Rydberg chain \cite{rader2019floating,PhysRevResearch.4.043102}. However, $\mathbb{Z}_p$ orders with odd $p$ in the ladder system break the reflection symmetry (see Fig.~\ref{fig:phasediag}\textbf{c} for the Rydberg density map of the $\mathbb{Z}_3$ order) and cannot form in the same way. If two $\mathbb{Z}_p$ chains are put together, the minimal energy case is to shift one chain by $\lfloor p/2\rfloor a$, leading to two Rydberg states occupying the diagonal sites in two rungs separated by $\lfloor p/2\rfloor a$. For odd $p$, $\mathbb{Z}_p$ orders require that there is a parameter regime where the Rydberg blockade radius $R_b$ is larger than $(p-1)a$ (to form $\mathbb{Z}_p$ order in each leg) but smaller than the diagonal in every two rungs separated by $(p-1)a/2$ (to allow both diagonal sites occupied by Rydberg states). In our system, the diagonal of NN rungs is $\sqrt{5}a > 2a$, so a large $\mathbb{Z}_3$ lobe shows up. However, the above conditions cannot be satisfied for $\mathbb{Z}_p$ orders with odd $p \ge 5$, precluding the existence of such density waves in the phase diagram. In the classical limit ($\Omega \rightarrow 0$), one can show that, due to interchain interactions, the energy of $\mathbb{Z}_p$ orders with odd $p\ge 5$ is always higher than other crystalline orders.

In $\mathbb{Z}_p$ ordered phase, there are $p$ degenerate states that break $\mathbb{Z}_p$ symmetry. Shifting one of these states by one lattice spacing $p-1$ times can generate the other degenerate states. This also applies to incommensurate floating phase where there are infinite degenerate states. In finite-size systems, different degenerate states have different boundary energy and the degeneracy is lifted. For Rydberg ladders with SBCs, the two extra sites on the boundary have a smaller Rydberg blockade effect due to a smaller number of neighbors and favor Rydberg excited states, which picks out one particular symmetry-breaking state that starts with Rydberg density $\langle n_1\rangle \approx 1$ and ends with $\langle n_{2L+2}\rangle \approx 1$. Notice that if the state with $\langle n_1 \rangle \approx 1$ chosen by the left boundary site is different from the one with $\langle n_{2L+2}\rangle\approx 1$ chosen by the right boundary site, domain walls will appear in the middle of the ladder. This puts a constraint on the length of the ladder to have clean crystalline orders in the system. To make the system size compatible with crystalline orders, the number of rungs of the ladder should be $L\equiv 0 \pmod{2}, 0 \pmod{3}, 1 \pmod{4}, 2 \pmod{6}$ for $\mathbb{Z}_2$, $\mathbb{Z}_3$, $\mathbb{Z}_4$, and $\mathbb{Z}_6$ orders, respectively. In our calculations for the phase diagram (Fig.~\ref{fig:phasediag}\textbf{b} and Fig.~\ref{fig:entphasediag}), we use $L = 288, 285$, and $290$ for $1 < R_b/a \le 2.5$ ($\mathbb{Z}_2$ and $\mathbb{Z}_3$), $2.5 < R_b/a \le 3.15$ ($\mathbb{Z}_3$ and $\mathbb{Z}_4$), and $3.15 < R_b/a \le 3.5$ ($\mathbb{Z}_6$), respectively. In the thermodynamic limit, the choice of different $L$s will not matter.

The peaks or the sudden jumps of entanglement entropy can portray all the phase boundaries. The boundary between the $\mathbb{Z}_2$ order and the disordered phase shows up as a line of peaks of $\mathcal{S}_{\rm{vN}}$, where the criticality is of Ising universality class with correlation length exponent $\nu=1$ and dynamical exponent $z=1$. The bright yellow line of peaks of $\mathcal{S}_{\rm{vN}}$ between the disordered phase and the floating phase is a Berezinskii-Kosterlitz-Thouless (BKT) transition line with an essential singularity in the correlation length as a function of parameters \cite{PhysRevLett.122.017205,Berezinsky:1970fr,Kosterlitz_1973,Kosterlitz_1974}. The quantum phase transition between the floating phase and crystalline orders is signaled by a sudden jump of $\mathcal{S}_{\rm{vN}}$ and is of Porkrovsky-Talapov (PT) universality class with $\nu = 1/2, z = 2$ \cite{Pokrovsky1979}. Looking at the melting of crystalline orders into the disordered phase, the $\mathbb{Z}_4$ order only directly shares a small portion of the boundary with the disordered phase (red solid line in Fig.~\ref{fig:entphasediag}), and shares the rest of the boundary with the floating phase. As crystalline orders melt, different types of domain-wall excitations that cost different energies yields chiral perturbations \cite{PhysRevB.24.5180,PhysRevB.24.398,HuseFisherPRL1982,huse1984commensurate}. If the chiral perturbation is relevant, the quantum phase transition between crystalline ordered and the disordered phases becomes chiral with $z>1$. Stronger chiral perturbations will lead to the critical floating phase with emergent U(1) symmetry between crystalline orders and the disordered phase. The critical exponents change continuously with parameters on the chiral transition line, which ends at Lifshitz points with $z=2$ where the floating phase bounded by BKT and PT lines appears \cite{PhysRevResearch.3.023049}. There should be a single CFT point with $z=1$ on the chiral transition line where the chiral perturbation vanishes, residing at the intersection of the commensurate line and the $\mathbb{Z}_4$ boundary \cite{Chepiga&Mila2021Kibble}. All of these predictions are observed around the $\mathbb{Z}_4$ order in Fig.~\ref{fig:entphasediag} and numerical evidence is provided in later sections. The Lifshitz points are found roughly around $(\Delta/\Omega, R_b/a) \approx (2.70, 2.65), (3.11, 2.825)$ by looking at where the floating phase starts to show up. The CFT point for $\mathbb{Z}_4$ order is accurately determined at $(\Delta/\Omega, R_b/a) = (2.7934, 2.7181)$ (see below for details). The same observation applies to $\mathbb{Z}_6$ order and the CFT point is found around $(3.327, 3.422)$.

The observations around the $\mathbb{Z}_4$ order in our ladder system are the same as those for $\mathbb{Z}_3$ and $\mathbb{Z}_4$ orders in 1D Rydberg chains, but the situations around $\mathbb{Z}_3$ and $\mathbb{Z}_6$ orders in the ladder system are very different. Unlike the 1D Rydberg chain, the $\mathbb{Z}_3$ order in the two-leg Rydberg ladder is fully covered by the floating phase and does not directly melt into the disordered phase. Notice that there is a narrow floating phase between the disordered phase and the $\mathbb{Z}_3$ order near the lower part of the $\mathbb{Z}_3$ lobe; for example, $(\Delta/\Omega, R_b/a) = (4, 2.1)$ is in the floating phase (see below). In the ladder, the $\mathbb{Z}_p$ state with odd $p$ breaks $\mathbb{Z}_2 \times \mathbb{Z}_p = \mathbb{Z}_{2p}$ symmetry, while the $\mathbb{Z}_p$ state with even $p$ still breaks $\mathbb{Z}_{p}$ symmetry. By shifting and flipping the ladder, there are six degenerate configurations for the $\mathbb{Z}_3$ order in the ladder system, while the degeneracy of the $\mathbb{Z}_3$ order in 1D chain is three. This means that according to the studies of clock models \cite{ORTIZ2012780,PhysRevB.100.094428,ZiQianLi2020}, continuous $U(1)$ symmetry can emerge from large discrete cyclic symmetry even without chiral perturbations. Therefore, the observation that the $\mathbb{Z}_3$ order is fully covered by a critical phase is reasonable. Similar explanations also apply to the 1D Rydberg chain, where there is always a critical phase between the $\mathbb{Z}_6$ order and the disordered phase due to strong chiral perturbations and the emergence of continuous symmetry from large discrete cyclic symmetry. However, in the two-leg Rydberg ladder, the strong interchain interaction may destabilize criticality so that direct melting of the $\mathbb{Z}_6$ order to the disordered phase is still possible (Fig.~\ref{fig:chiralcftz6}). In the following sections, we provide substantial numerical evidence for all the above analyses about the phase diagram in Fig.~\ref{fig:entphasediag}.

\begin{figure*}[t]
\centering
\includegraphics[width=1\textwidth]{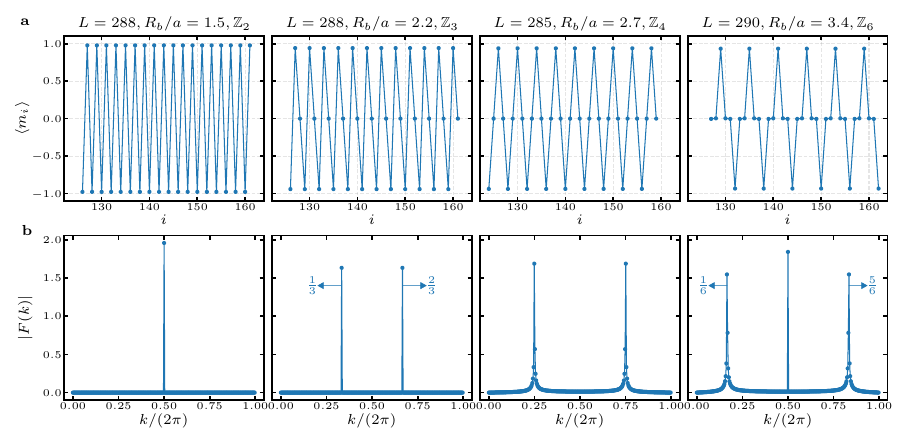}
\caption{\textbf{The profiles of density difference $m_i = n_{i,1}-n_{i,0}$ for crystalline orders and the corresponding Fourier transforms.} \textbf{a}, The four plots are for $\mathbb{Z}_2$, $\mathbb{Z}_3$, $\mathbb{Z}_4$, and $\mathbb{Z}_6$ orders along $\Delta/\Omega=4$ cut (right-pointing triangles in Fig.~\ref{fig:entphasediag}), respectively. \textbf{b}, Absolute values of the discrete Fourier transform output as a function of the wave vector $k$.}
\label{fig:mifmiorder}
\end{figure*}
\begin{figure*}[t]
\centering
\includegraphics[width=1\textwidth]{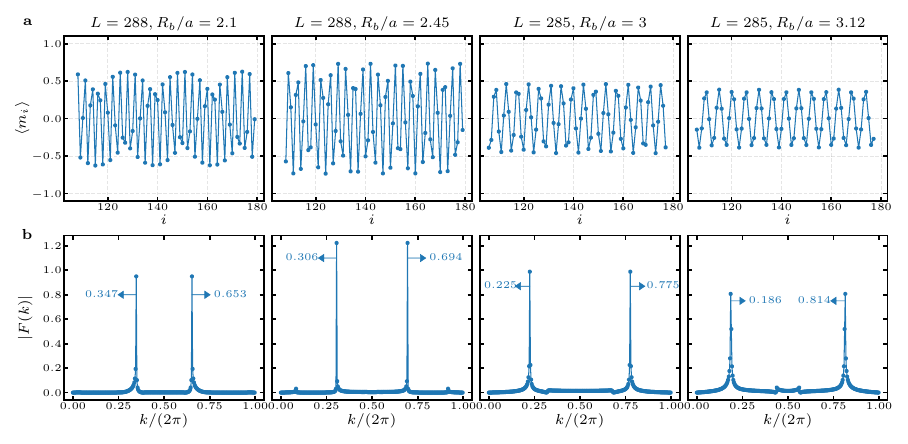}
\caption{\textbf{Analogous plots as Fig.~\ref{fig:mifmiorder}, but for the floating phase.} The four columns show the results for the floating phase with a wave vector between $2\pi/2$ and $2\pi/3$, $2\pi/3$ and $2\pi/4$, $2\pi/4$ and $2\pi/5$, and $2\pi/5$ and $2\pi/6$ along $\Delta/\Omega=4$ cut (down-pointing triangles in Fig.~\ref{fig:entphasediag}), respectively.}
\label{fig:mifmifloating}
\end{figure*}

\begin{figure*}[t]
\centering
\includegraphics[width=1\textwidth]{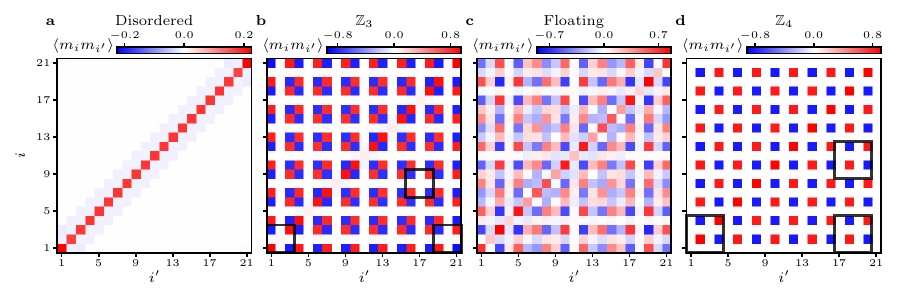}
\caption{\textbf{The numerical correlation matrix $\langle m_i m_{i'} \rangle$ employing the order parameter $m_i = n_{i,1}-n_{i,0}$ for a system size of $L=21$, in comparison with the experimental results in Fig.~\ref{fig:corrsl21}\textbf{a} of the main text.} The presented results correspond to $(\Delta/\Omega, R_b/a) = (0, 2.4)$ in the disordered phase (\textbf{a}), $(3.5, 2.1)$ in the $\mathbb{Z}_3$ order (\textbf{b}), $(3.5, 2.45)$ in the floating phase (\textbf{c}), and $(3.5, 2.7)$ in the $\mathbb{Z}_4$ order (\textbf{d}). Black boxes indicate examples of minimal repeating patterns for $\mathbb{Z}_3$ and $\mathbb{Z}_4$ orders, which repeat every three or four sites, respectively. The quantum floating phase lacks a repeating pattern due to an incommensurate wavelength. }
    \label{fig:numcorrmatrx}
\end{figure*}

\begin{figure*}[t]
\centering
\includegraphics[width=1\textwidth]{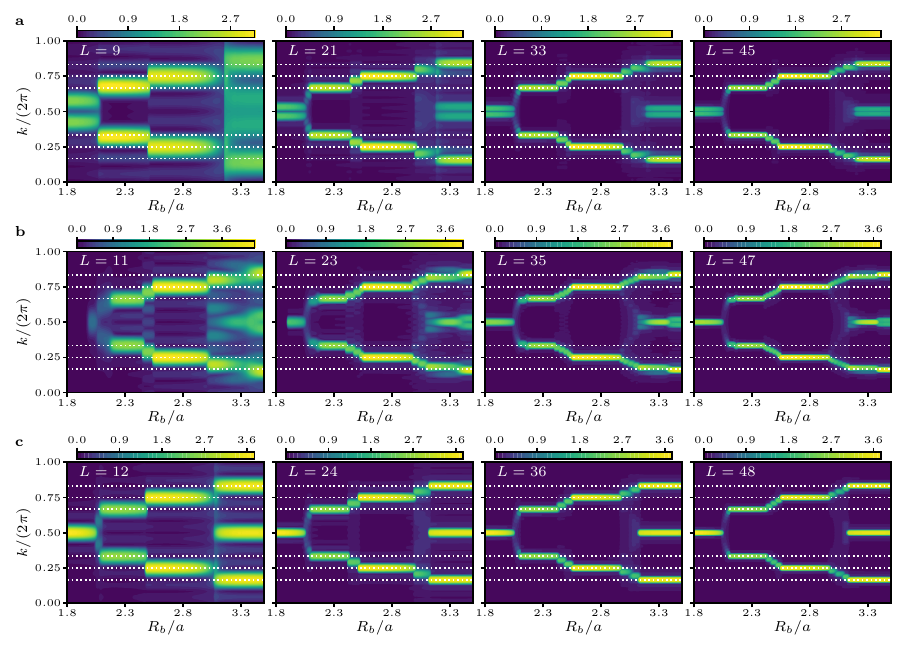}
\caption{\textbf{The effects of boundary conditions on the structure factor $S(k)$.} The results are for the vertical cut $\Delta/\Omega = 4$. \textbf{a}, The four plots in the first row are for SBCs, where the boundary sites favor Rydberg states. The system sizes with $L \equiv 9 \pmod{12}$ are compatible with both $\mathbb{Z}_3$ and $\mathbb{Z}_4$ orders, but not with $\mathbb{Z}_2$ or $\mathbb{Z}_6$ orders. \textbf{b}, For OBCs, the system sizes with $L\equiv 11 \pmod{12}$ are compatible with both $\mathbb{Z}_3$ and $\mathbb{Z}_4$ orders. They are also compatible with $\mathbb{Z}_2$ order due to weak constraints of Rydberg density on the boundaries, but not compatible with $\mathbb{Z}_6$ order. \textbf{c}, For PBCs, the system sizes with $L\equiv 0 \pmod{12}$ are compatible with all $\mathbb{Z}_2$, $\mathbb{Z}_3$, $\mathbb{Z}_4$, and $\mathbb{Z}_6$ orders. }
\label{fig:strucfacfor3bcs}
\end{figure*}
\begin{figure}
\centering
\includegraphics[width=\linewidth]{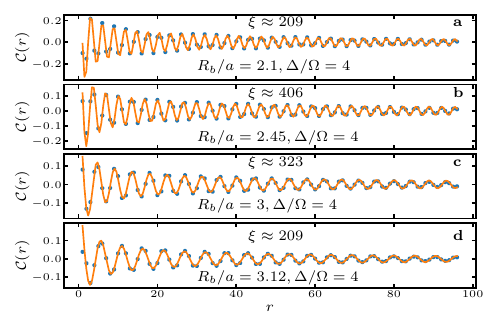}
\caption{\textbf{Numerical calculations of the correlation lengths $\xi$ for the floating phase.} The results are obtained from DMRG for $L=288$ and the parameters are the same as Fig.~\ref{fig:mifmifloating}. The two-point correlation functions $\mathcal{C}(r) = \langle m_{L/2}m_{L/2+r} \rangle - \langle m_{L/2}\rangle \langle m_{L/2+r}\rangle$ ($r=1,2,\ldots,96$) are presented and fit to the Ornstein–Zernicke form ($r=3,4,\ldots,96$).}
\label{fig:corrlengthsfloating}
\end{figure}
\begin{figure*}[t]
\centering
\includegraphics[width=1\textwidth]{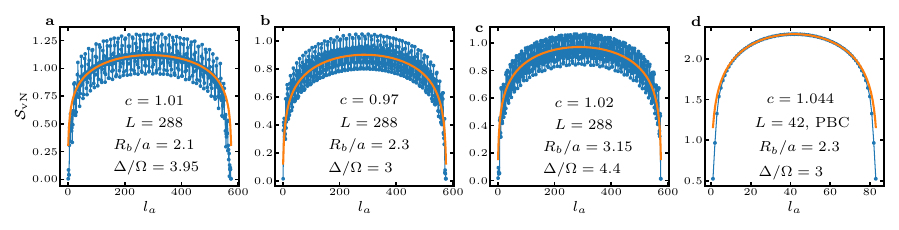}
\caption{\textbf{The dependence of the entanglement entropy $\mathcal{S}_{\rm{vN}}$ on the subsystem size $l_a$ for the floating phase.} $L$ is the number of rungs of the ladder and $l_a$ is the number of sites of the subsystem. The curve fit of $\mathcal{S}_{\rm{vN}}$ to the CFT form is performed using the middle $2/3$ of values of $l_a$. The results for the floating phases between $\mathbb{Z}_2$ and $\mathbb{Z}_3$ orders, $\mathbb{Z}_3$ and $\mathbb{Z}_4$ orders, and $\mathbb{Z}_4$ and $\mathbb{Z}_6$ orders are presented in \textbf{a}, \textbf{b}, and \textbf{c} (red left-pointing triangles in Fig.~\ref{fig:entphasediag}), respectively. \textbf{d}, The results for the system with $L=42$ and PBC are presented.}
\label{fig:floatingEEfit}
\end{figure*}

\begin{figure*}[t]
\centering
\includegraphics[width=1\textwidth]{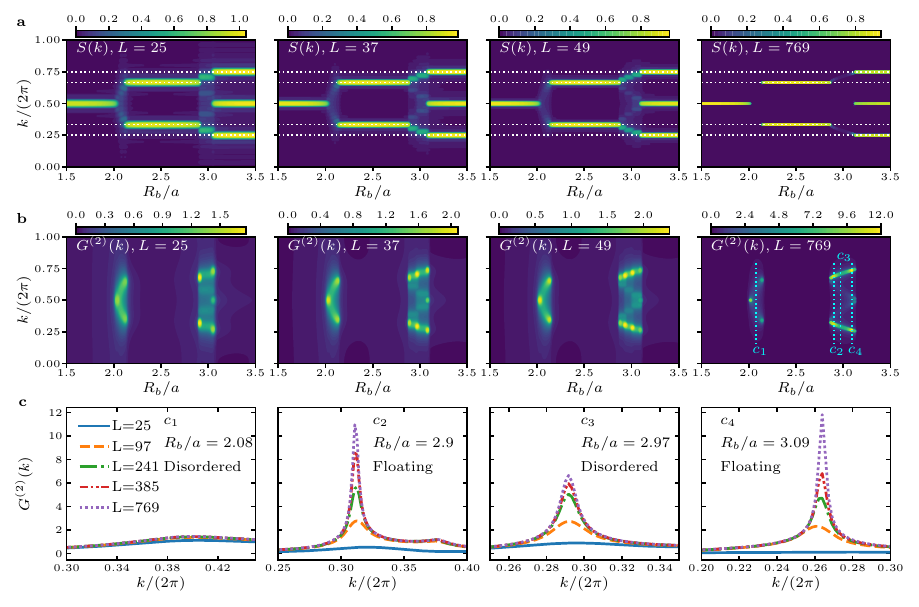}
\caption{\textbf{Structure factors $S(k)$ and connected density-density correlators $G^{(2)}(k)$ for a single-chain Rydberg array.} The results are for the vertical cut $\Delta/\Omega=3$ of the phase diagram shown in Ref.~\cite{PhysRevResearch.4.043102}. \textbf{a}, The plots of $S(k)$ for $L=25, 37, 49, 769$ are presented in the first row. The white dotted lines label $k=2\pi/4, 2\pi/3, 4\pi/3$, and $6\pi/4$, respectively. \textbf{b}, $G^{(2)}(k)$ is plotted for the same system sizes. \textbf{c}, The dependence of $G^{(2)}(k)$ on the system sizes for the disordered phase and the floating phase is presented for vertical cuts labeled in the fourth plot in \textbf{b}.}
\label{fig:strucfacforonechain}
\end{figure*}

\begin{figure}
\centering
\includegraphics[width=\linewidth]{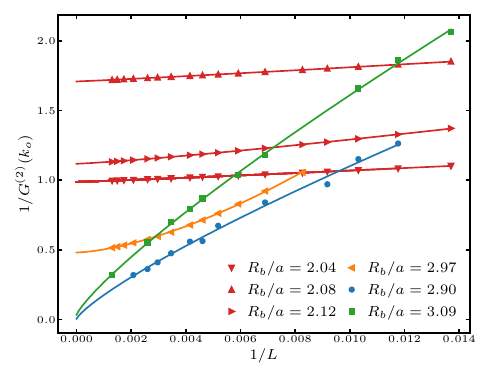}
\caption{\textbf{Finite-size scaling of the inverse peak height of $G^{(2)}(k_o)$ for the 1D Rydberg chain.} The results are for the $\Delta/\Omega=3$ cut. For $R_b/a = 2.04, 2.08, 2.12$, the system is in the disordered phase between $\mathbb{Z}_2$ and $\mathbb{Z}_3$ orders. $R_b/a = 2.9$ is inside the floating phase near $\mathbb{Z}_3$ order, and $R_b/a = 3$ is inside the floating phase near $\mathbb{Z}_4$ order, as indicated by the divergence of $G^{(2)}(k_o)$ in the thermodynamic limit. For $R_b/a = 2.97$, the system is in the disordered phase between the above two floating phases, as indicated by the convergence of $G^{(2)}(k_o)$ to a finite value in the thermodynamic limit.}
\label{fig:g2peakscaling1chain}
\end{figure}

\begin{figure*}[t]
\centering
\includegraphics[width=1\textwidth]{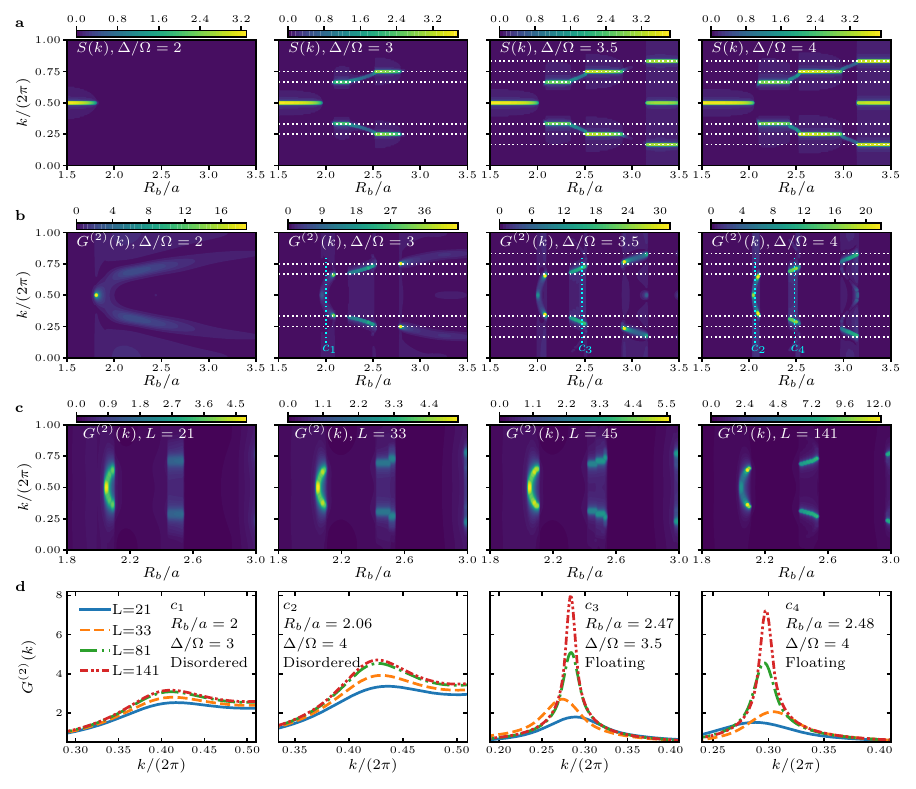}
\caption{\textbf{Structure factors $S(k)$ and connected density-density correlators $G^{(2)}(k)$ for the two-leg Rydberg ladder.} \textbf{a}, The four plots in the first row are for four vertical cuts $\Delta/\Omega=2, 3, 3.5, 4$ using the date set of the phase diagram in Fig.~\ref{fig:phasediag}\textbf{b}. \textbf{b}, The same as \textbf{a}, but for $G^{(2)}(k)$. \textbf{c}, $G^{(2)}(k)$ for different system sizes $L=21, 33, 45, 141$ are plotted at fixed $\Delta/\Omega = 4$ and $1.8 \le R_b/a \le 3$. \textbf{d}, The dependence of $G^{(2)}(k)$ on the system sizes for the disordered phase and the floating phase is presented for the four cuts labeled in \textbf{b}.}
\label{fig:strucfacfortwochains}
\end{figure*}
\begin{figure}[t]
\centering
\includegraphics[width=\linewidth]{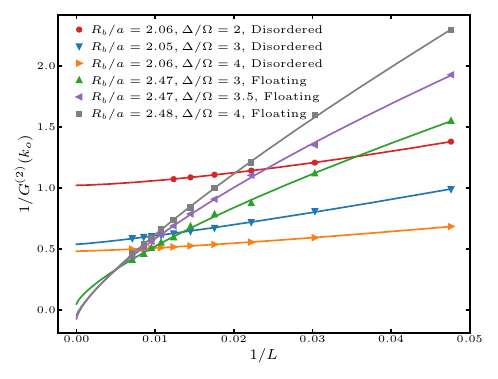}
\caption{\textbf{Finite-size scaling of the inverse peak height of $G^{(2)}(k_o)$ for the two-leg Rydberg ladder.} Three cases in the disordered phase and three cases in the floating phase between $\mathbb{Z}_3$ and $\mathbb{Z}_4$ orders are presented (magenta up-pointing triangles in Fig.~\ref{fig:entphasediag}).}
\label{fig:g2peakscaling2chains}
\end{figure}
\begin{figure*}[t]
\centering
\includegraphics[width=1\textwidth]{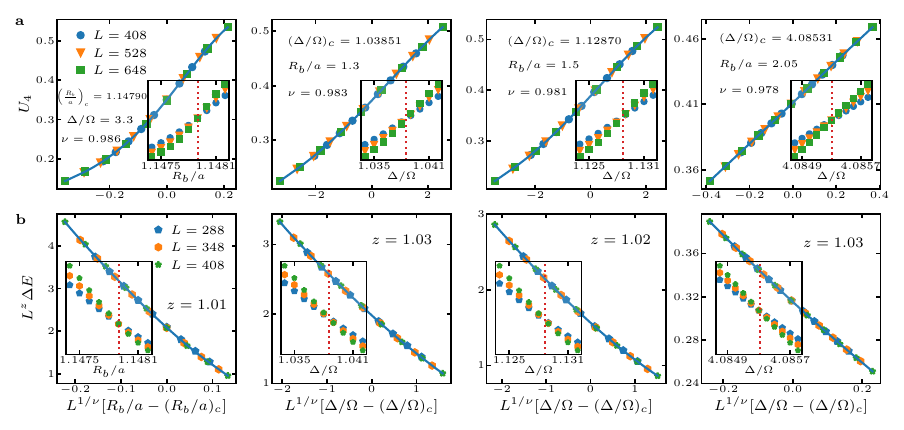}
\caption{\textbf{Data collapse of the Binder Cumulant $U_4$ and the rescaled energy gap $L^z\Delta E$ for quantum phase transitions between the $\mathbb{Z}_2$ order and the disordered phases.} The results are computed using DMRG for the white circles labeled on the $\mathbb{Z}_2$ boundary of the phase diagram in Fig.~\ref{fig:entphasediag}. \textbf{a}, The critical point and the correlation length exponent $\nu$ are fine tuned to find the best data collapse of $U_4$ for $L=408, 528, 648$. The inset shows the same data set as a function of the original parameter. The first plot is for the phase transition along the $\Delta/\Omega=3.3$ cut and the data collapses onto a universal function of $L^{1/\nu}[R_b/a-(R_b/a)_c]$. The other three plots are for phase transitions along $R_b/a=1.3, 1.5, 2.05$ cuts, respectively, and the data collapse onto universal functions of $L^{1/\nu}[\Delta/\Omega-(\Delta/\Omega)_c]$. \textbf{b}, Using the values of critical points and $\nu$ obtained in \textbf{a}, the dynamical exponent $z$ is determined by finding the best data collapse of $L^z\Delta E$ for $L=288, 348, 408$. The inset shows the same data set as a function of the original parameter.}
\label{fig:collapseu4gapising}
\end{figure*}

\section{Density Profiles, Structure Factors, and Effects of Boundary Conditions}

\subsection{Numerical Density Profiles and Fourier Transforms}

We first show numerical results for the density profiles and corresponding Fourier transforms for the crystalline orders (Fig.~\ref{fig:mifmiorder}) and the floating phase (Fig.~\ref{fig:mifmifloating}) for a large system size. For a two-leg ladder, the order parameter is the Fourier transform of the local density, $n(k_x; k_y=0,\pi) \sim \sum_{i,j} \exp(\mathrm{i}k_x i + \mathrm{i}k_y j) \langle n_{i,j}\rangle$ where $i=1,2,\ldots,L$ is the rung index and $j=0,1$ is the leg index. In our case, all the density-wave orders in the phase diagram break the up-down $\mathbb{Z}_2$ symmetry and $\langle n_{i,1}-n_{i,0}\rangle$ is finite. Thus, it is enough to probe those orders only using $k_y=\pi$ in the Fourier transform. In the following, we will omit the subscript and just use $k$ in the Fourier transform. The profiles for the density difference $\langle m_i \rangle = \langle n_{i,1}-n_{i,0}\rangle$ of each rung are plotted in Fig.~\ref{fig:mifmiorder}\textbf{a} for crystalline orders. For $\mathbb{Z}_2$, $\mathbb{Z}_4$, and $\mathbb{Z}_6$ orders, the profile of $m_i$ takes the form ``$\cdots,+1,-1,+1,-1,\cdots$'', ``$\cdots+1,0,-1,+1,0,-1,\cdots$'', and ``$\cdots+1,0,0,-1,0,0,+1,0,0,-1,0,0,\cdots$'', respectively. The degeneracy of the symmetry breaking states is equal to the periodicity of the density wave. The profile of the $\mathbb{Z}_3$ order takes the form ``$\cdots,+1,-1,0,+1,-1,0,\cdots$'', and there are $6$ degenerate states. Notice that the maximal value of $|\langle m_i\rangle|$ should be smaller than $1$ in the quantum regime. The absolute value of the Fast Fourier Transform (FFT) of $\langle m_i\rangle$, $|F(k)|=|p/L\sum_i \exp(\mathrm{i}ki) \langle m_i\rangle|$ ($p$ is the wavelength of the density wave), is presented in Fig.~\ref{fig:mifmiorder}\textbf{b}. It is seen that the peak of $|F(k)|$ is located at $2\pi/p$ and $2\pi-2\pi/p$ as expected. For the $\mathbb{Z}_6$ order, $|F(k)|$ also has a peak at $\pi$, which is easy to check using the bitstring of $m_i$. Similarly, the profile of $\langle m_i\rangle$ for the floating phase is presented in Fig.~\ref{fig:mifmifloating}\textbf{a}. We show four floating phases that have fractional wavelengths between $2$ and $3$, between $3$ and $4$, between $4$ and $5$, and between $5$ and $6$, respectively. The FFT in Fig.~\ref{fig:mifmifloating}\textbf{b} also gives the expected wave vector. For the floating phase at $\Delta/\Omega=4, R_b/a=2.1$, the wave length is very close to integer $3$ and the wave vector $k_o$ is close to $2\pi/3$. A density wave like $\cos(k_oi) = \cos[(k_o-2\pi/3)i+2\pi i/3] \approx \cos[(k_o-2\pi/3)i]\cos(2\pi i/3)$, so the small frequency difference $|k_o-2\pi/3|$ gives rise to a big wave packet in the profile.

\subsection{Structure Factors}
Local density profiles of density wave orders can be mixed with domain walls due to finite size and boundary effects, and the Friedel oscillations from the boundary may overwhelm the signal of the bulk oscillations. The correlation functions are more suitable to study the bulk properties associated with quantum phase transitions. We have experimentally measured the structure factor $S(k)$, which is the Fourier transform of density-density correlations $\langle m_i m_{i'} \rangle$ as defined in Eq.~\eqref{eq:strucfactor} in the main text. The numerical density-density correlations $\langle m_i m_{i'} \rangle$ are presented in Fig.~\ref{fig:numcorrmatrx}. Same as the experimental results, the correlation matrix for the disordered phase has non-zero diagonal elements and nearly zero off-diagonal elements, but the numerical off-diagonal elements have no site-dependent fluctuations, which the experimental results have as shown in the main text. In the crystalline orders, the numerical correlations show perfect periodic oscillations without any amplitude decay with the distance $|i-i'|$, demonstrating the existence of long-range orders. The black boxes in Figs.~\ref{fig:numcorrmatrx}\textbf{b} and \textbf{d} encircle the smallest repeating units for $\mathbb{Z}_3$ and $\mathbb{Z}_4$ orders, respectively. No such repeating unit is observed in the numerical correlation matrix for the floating phase, indicating that the floating phase has a wavelength that is incommensurate with the lattice spacing.

In our experiment, one of the limitations of creating ordered phases is the introduction of errors due to quantum fluctuations and a thermal environment. This is evidenced by a finite correlation length, $\xi$, as an exponential amplitude decay $e^{-|i-i'|/\xi}$ between correlations across two points $(i,i')$ during single shot measurements and averaged across the whole experiment.

\subsection{Effects of Different Boundary Conditions}

In Fig.~\ref{fig:strucfacfor3bcs}, we show more numerical results on the structure factor in small systems along $\Delta/\Omega = 4$ cut and illustrate the boundary effects. For SBCs, we show the results for four system sizes $L=9,21,33,45\ [\equiv 9 \pmod{12}]$ that are compatible with both $\mathbb{Z}_3$ and $\mathbb{Z}_4$ orders, but not with $\mathbb{Z}_2$ and $\mathbb{Z}_6$ orders. In the regime of $\mathbb{Z}_3$ and $\mathbb{Z}_4$ orders, the structure factor shows the correct peaks at $k=2\pi/3$ and $2\pi/4$. But in the regime of $\mathbb{Z}_2$ and $\mathbb{Z}_6$ orders, the peak at $k=\pi$ is split into two and the peak at $k=2\pi/6$ is shifted due to domain wall defects. Using SBCs, the states should have inversion symmetry around the center of the ladder such that the sum in $S(k=\pi)$ are almost canceled and the peak at $\pi$ splits. These effects diminish as $L$ increases.

For OBCs, the boundary rungs also favor Rydberg states, the four system sizes presented are $L=11,23,35,47\ [\equiv 11 \pmod{12}]$, which are also compatible with $\mathbb{Z}_3$ and $\mathbb{Z}_4$ orders. For $L=11$, most regime of the $\mathbb{Z}_2$ order has no up-down symmetry breaking in DMRG calculations and there is no peak in $S(k)$, while the symmetry breaking happens for larger system sizes. For $R_b/a<2$, the size of rungs is larger than the blockade radius, the boundary rungs can have double occupations such that the bulk has a clean $\mathbb{Z}_2$ order. The peak at $\pi$ does not split into two, so the system sizes $L\equiv 11 \pmod{12}$ are also compatible with $\mathbb{Z}_2$ order. However, $L\equiv 11 \pmod{12}$ is not compatible with $\mathbb{Z}_6$ order, so the peak at $2\pi/6$ is shifted. The small jump of the plateau around $R_b/a=3.4$ comes from the change of symmetry of the state from reflection symmetry to inversion symmetry (signaled by the splitting of the $\pi$ peak). These discrepancies observed above again diminish as $L$ increases. Finally, the systems with $L=12,24,36,48\ [\equiv 0 \pmod{12}]$ and PBCs are compatible with all $\mathbb{Z}_p$ ($p=2,3,4,6$) orders and the peak of $S(k)$ precisely resides at $2\pi/p$.

Using the proper normalization factor described in the main text, the peak height of $S(k)$ is $S(2\pi/p) = 4$ for the classical $\mathbb{Z}_p$ orders with even $p$, and $S(2\pi/p) = 2+2\cos(\pi/p)$ for odd $p$.  In the quantum regime, the peak height is lower than that in the classical limit. The peak heights of $S(k)$ in Fig.~\ref{fig:strucfacfor3bcs} and Fig.~\ref{fig:strucfacfortwochains} are consistent with the above expectations. Assuming that the local density for the floating phase is $\langle m_i\rangle \sim \cos(k_fi)$, where $k_f = 2\pi/p_f$ is the incommensurate wave vector, one can show that at least $S(k_f) \sim p_f^2/4$ for the floating phase in the large $L$ limit, while $S(k)$ is zero in the disordered phase. These facts can be used to differentiate between the floating phase and the disordered phase in the incommensurate regime. 

In the incommensurate regime between crystalline orders, the wave vectors change continuously with the physical parameters in the thermodynamic limit. For finite systems, however, the constraints on the amplitude of density waves on the boundaries discretize the allowed wave vectors. For SBCs, the Rydberg density on boundary sites tends to be $1$ and only certain wavelengths can satisfy the boundary conditions. The PBCs also naturally discretize the wave vectors. If $L=12l-3$ for SBCs or $L=12l$ ($l$ is an integer) for PBCs, the discrete wave vectors allowed in the floating phase takes $2l-1$ values between $\mathbb{Z}_2$ and $\mathbb{Z}_3$ orders, $l-1$ values between $\mathbb{Z}_3$ and $\mathbb{Z}_4$ orders, and $l-1$ values between $\mathbb{Z}_4$ and $\mathbb{Z}_6$ orders. For OBCs, the boundary rungs favor Rydberg states that can reside in the same leg or different legs. The constraints on the density waves for OBCs are weaker than those for SBCs and PBCs. There are more allowed wave vectors between crystalline orders. The results in Fig.~\ref{fig:strucfacfor3bcs} confirm all the above expectations. Notice that the incommensurate regime between $\mathbb{Z}_2$ and $\mathbb{Z}_3$ orders is mainly the disordered phase, so most of the peak plateaus of $S(k)$ decay as $L$ increases. On the other hand, the narrow peak plateaus that do not decay with increasing $L$ near $\mathbb{Z}_3$ plateau demonstrate the existence of a narrow floating phase there. In the thermodynamic limit, $l \rightarrow \infty$, the wave vector changes continuously with the physical parameters. 

A Rydberg ladder with PBCs is a three dimensional cylinder with a height $a_y$ and is not realizable in our experiments. For OBCs, because the geometry is symmetric under flipping the leg index, the adiabatic time evolution initialized from all atoms in the $\ket{g}$ state will randomly prepare one of two degenerate states related by flipping the leg index, which obfuscates the observation of the density waves from the average density profile. Figure~\ref{fig:strucfacfor3bcs} shows that the structure factor for SBCs has cleaner signals than that for OBCs in our experimental regime from the $\mathbb{Z}_3$ order to the $\mathbb{Z}_4$ order, so we use SBCs in our experiments.

\section{Criticality of the Quantum Floating Phase}
Based on symmetry arguments, the emerging incommensurate quantum floating phase with continuous U(1) symmetry can be explained by simple chiral $\mathbb{Z}_p$ clock models \cite{RhineChiralZ32018,NYCKEES2021115365}. In fact, there is a simple intuitive picture for the incommensurate density waves with an irrational wave length (in units of lattice spacing $a$). Assuming the density wave is described by $\cos(k_{f}i)$ with $2\pi/k_{f}$ being an irrational number, the phase $k_{f}i \pmod{2\pi}$ takes different values for any two different $i$s and the local density profile never repeats. When $i$ takes all integer values, $k_{f}i \pmod{2\pi}$ exhausts all values between $0$ and $2\pi$. So an arbitrary shift in the phase does not change the incommensurate wave and the continuous translational symmetry emerges. Here, the periodicity is infinite and $\mathbb{Z}_{p}$ symmetry with $p \rightarrow \infty$ becomes a U(1) symmetry. Theoretically, the floating phase should be a critical phase that has infinite degeneracies and it can be described by the Tomonaga-Luttinger Liquid (TLL) effective theory \cite{F.D.M.Haldane_1981,PhysRevB.84.085114}. In this section below, we provide numerical evidence to show the criticality of the quantum floating phase by the divergence of the correlation length and find that its central charge $c=1$ by fitting the entanglement entropy.

\subsection{Correlation Length}
To demonstrate that the floating phase is critical, we first show numerically that the correlation length diverges. The connected density-density correlators $\mathcal{C}(r) = \langle m_{L/2}m_{L/2+r} \rangle - \langle m_{L/2}\rangle \langle m_{L/2+r}\rangle$ ($r=1,2,\ldots,96$) for $L=288$ are presented in Fig.~\ref{fig:corrlengthsfloating} based on DMRG calculations, where the four floating phases have the same parameters as those in Fig.~\ref{fig:mifmifloating}. One can see that $\mathcal{C}(r)$ decreases slowly with the distance $r$ for all the floating phases shown here, indicating that the correlation length is large. We fit the connected correlators to the Ornstein–Zernicke form \cite{OrnsteinCorr}
\begin{eqnarray}
\label{eq:corrozform}
\mathcal{C}(r) \sim \frac{e^{-r/\xi}}{\sqrt{r}} \cos(kr+\phi_0)
\end{eqnarray}
and extract the correlation length $\xi$. The first two data points for $r=1,2$ are discarded in the curve fit. It can be seen that the quality of curve fit is excellent for all the cases and the extracted $\xi$ is $209$, $406$, $323$, and $209$ for $R_b/a = 2.1$, $2.45$, $3$, and $3.12$ on the $\Delta/\Omega=4$ cut, respectively. All the fitted values of $\xi$ are of the same order as the system size $L=288$, indicating that the correlation length diverges in the thermodynamic limit and the floating phase is gapless.

\subsection{Entanglement Entropy}
Another feature of the critical phase in 1D system is that the entanglement entropy as a function of subsystem size should satisfy the formula in Eq.~\eqref{eq:ent}. Here, the periodic oscillations of density waves should also result in fluctuations in $\mathcal{S}_{\rm{vN}}$ on top of Eq.~\eqref{eq:ent}. We plot $\mathcal{S}_{\rm{vN}}$ as a function of the number of sites of the subsystem $l_a$ for three floating phases between $\mathbb{Z}_2$ and $\mathbb{Z}_3$ orders, $\mathbb{Z}_3$ and $\mathbb{Z}_4$ orders, and $\mathbb{Z}_4$ and $\mathbb{Z}_6$ orders in Fig.~\ref{fig:floatingEEfit}\textbf{a}, \textbf{b}, and \textbf{c}, respectively. The number of rungs for the ladder system is $L=288$ for the first three subplots. In Fig.~\ref{fig:floatingEEfit}\textbf{d}, the results for the floating phase between $\mathbb{Z}_3$ and $\mathbb{Z}_4$ orders with $L=42$ and PBC (the truncation error $\epsilon=10^{-9}$ is used in DMRG) are presented. We remark that the crystalline orders have short-range correlations and are gapped phases, area laws \cite{RevModPhys.82.277} ensure that the entanglement entropy of $\mathbb{Z}_p$ symmetry breaking phases is close to zero and does not scale with the system size $L$ (not shown here).

It can be seen that there indeed exist oscillations in $\mathcal{S}_{\rm{vN}}$ for the floating phase with SBCs. The pattern of the oscillations is not clear, but Eq.~\eqref{eq:ent} should correctly describe the main behavior of $\mathcal{S}_{\rm{vN}}$. We fit the formula in Eq.~\eqref{eq:ent} to $\mathcal{S}_{\rm{vN}}$, where $1/6$ of the data points on both sides are discarded to reduce the boundary effects. The fitted values of the central charge are $c=1.01$, $0.97$, and $1.02$ for the three finite-size floating phases with SBCs, respectively. For the system with $L=42$ and PBC, there is no oscillation in $\mathcal{S}_{\rm{vN}}$. Due to the translational invariance in units of lattice spacing in systems with PBCs and the incommensurate wave vector of the floating phase, translational symmetry breaking cannot happen and the density wave form of the floating phase does not show up. The formula in Eq.~\eqref{eq:ent} for PBCs applies to the entanglement entropy in Fig.~\ref{fig:floatingEEfit}\textbf{d} and the fitted central charge is $c=1.044$. All the results are consistent with the expectation that the central charge of the floating phase should be $c=1$, which means the low-energy effective theory is described by TLL \cite{NishimotoTLLEnt2011} or the sine-Gordon model \cite{ColemanSG1975,LECHEMINANT2002502}.
\section{1D Rydberg Chain versus Two-leg Rydberg Ladder}

In this section, we numerically compare and contrast the 1D Rydberg chain setup with our two-leg Rydberg ladder setup and show that the floating phase regions are much broader in our two-leg ladder array than the 1D chain.  First, we introduce the connected density-density correlator and demonstrate subsequently that the conclusions drawn from the connected density-density correlators are consistent with our results obtained from the structure factors. We examine the 1D Rydberg chain and show that the connected density-density correlators diverge in the thermodynamic limit at the peak wave vector in the floating phase. There are narrow regions of quantum floating phase between the $\mathbb{Z}_3$ and $\mathbb{Z}_4$ orders, but most of the incommensurate regime there is in the disordered phase. These conclusions are consistent with previous studies~\cite{PhysRevLett.122.017205, rader2019floating, Chepiga&Mila2021Kibble, PhysRevResearch.3.023049, PhysRevResearch.4.043102}. 

On the other hand, similar analysis shows that there are broad regions of the quantum floating phase between the $\mathbb{Z}_3$ and $\mathbb{Z}_4$ orders for the two-leg ladder system. As is analysed before, interactions among the two legs of the ladder in the chosen geometric aspect ratio can lead to the appearance of the $\mathbb{Z}_p$ translational-symmetry-breaking states at much smaller $R_b/a$. In particular, the $\mathbb{Z}_3$ state in our two-leg ladder, which breaks the $\mathbb{Z}_2\times \mathbb{Z}_3$ symmetry, has more types of domain walls in the melting process than the $\mathbb{Z}_3$ state in the 1D Rydberg chain does. Intuitively, we suspect that more types of domain walls can generate stronger chiral perturbations between the crystalline ordered and the disordered phases, which in turn produce bigger regions of the floating phase. The existence of broad regions of the floating phase in experimentally accessible parameters of $R_b/a$ greatly facilitate our experimental observation of the floating phase. 

\subsection{Connected Density-Density Correlator}

We have numerically demonstrated in the previous section that the incommensurate regime with high entanglement plateau in the phase diagram is a critical floating phase. The connected density-density correlation function $\langle m_i m_{i'} \rangle -\langle m_i \rangle \langle m_{i'} \rangle$ quantifies the extent of quantum fluctuations in real space, and its Fourier transform
\begin{eqnarray}
\label{eq:connectedcorr}
G^{(2)}(k) = \frac{p}{L} \sum_{i,i'} e^{\mathrm{i}k \left(i-i' \right)} \left(\langle m_i m_{i'} \rangle -\langle m_i \rangle \langle m_{i'} \rangle \right)
\end{eqnarray}
is an analog of magnetic susceptibility and diverges in the critical floating phase at the density wave vector $k_o$ (the peak position of the structure factor). Here, we also use the effective length $L/p$ as the normalization factor, where $p=2\pi/k_o$ is the wave length of the density wave and can be determined by the peak position of $S(k)$. In crystalline orders, the correlation is short-range and $G^{(2)}(k)$ is close to zero. As long as the system is in a gapped phase and the correlation length is finite, the peak height of $G^{(2)}(k)$ would be finite.

\subsection{One-dimensional Rydberg Chain}

In order to check the above statements, we first examine the results for the well studied 1D Rydberg chain. In Fig.~\ref{fig:strucfacforonechain}, we show the structure factor $S(k)$ and the connected density-density correlator $G^{(2)}(k)$ for the 1D Rydberg chain on the $\Delta/\Omega=3$ cut. The full phase diagram of the 1D chain can be found in Ref.~\cite{PhysRevResearch.4.043102}. We focus on the regime $1.5 < R_b/a < 3.5$ where $\mathbb{Z}_2$, $\mathbb{Z}_3$, and $\mathbb{Z}_4$ orders appear successively. The system sizes (number of sites) with $L \equiv 1 \pmod{12}$ are compatible with all three crystalline orders. Figure~\ref{fig:strucfacforonechain}\textbf{a} shows the structure factor $S(k)$ as a function of $k$ and $R_b/a$ defined in the same way as Eq.~\eqref{eq:strucfactor} in the main text. The mean density $\bar{n}=(\sum_i \langle n_i \rangle)/L$ is subtracted from the local density in the calculation to remove the peak of $S(k)$ at $k=0$. For classical $\mathbb{Z}_p$ orders, one can show that the structure factor for the 1D Rydberg chain is always $1$ at $k=2\pi/p$ and $2\pi-2\pi/p$. We can see that the numerical values of $S(k)$ for $L=25, 37, 49, 769$ precisely depict the peak at $k=2\pi/p$ and $2\pi-2\pi/p$ with peak height close to $1$. It can also be seen that the peak width decreases with increasing $L$. The structure factor in the incommensurate regime between $\mathbb{Z}_2$ and $\mathbb{Z}_3$ orders does not show obvious peaks even for the small system with $L=25$ and decreases close to zero as $L$ increases. On the other hand, there are clear peaks in $S(k)$ between $\mathbb{Z}_3$ and $\mathbb{Z}_4$ orders. Because the boundary sites favor Rydberg excited states, particular density waves starting with $n_1\approx 1$ and ending with $n_L\approx 1$ are chosen and the allowed wave vectors are discretized. For $L=12l+1$ ($l$ is an integer), the allowed wave vectors in units of $2\pi$ takes $l-1$ values between $\mathbb{Z}_3$ and $\mathbb{Z}_4$ orders: $(3l+1)/(12l), (3l+2)/(12l), \ldots, (4l-1)/(12l)$. The results for $L=25,37,49$ in Fig.~\ref{fig:strucfacforonechain}\textbf{a} confirm the above arguments. These peaks suggest that there exist an floating phase between $\mathbb{Z}_3$ and $\mathbb{Z}_4$ orders. For very large $L = 769$, the peak height of $S(k)$ for most of the regime between $\mathbb{Z}_3$ and $\mathbb{Z}_4$ orders becomes close to zero, while the peak height in very narrow parts near the crystalline boundaries remains finite. This indicates that there only exists a narrow floating phase near the boundaries of crystalline orders and most of the incommensurate regime between $\mathbb{Z}_3$ and $\mathbb{Z}_4$ orders is in the disordered phase.

The information about the critical floating phase can also be read out from $G^{(2)}(k)$ in Fig.~\ref{fig:strucfacforonechain}\textbf{b}. In the regime of crystalline orders, $G^{(2)}(k)$ is nearly zero, while it is large in the incommensurate regime between crystalline orders. It is clear that the regime between $\mathbb{Z}_2$ and $\mathbb{Z}_3$ orders is the disordered phase, where $G^{(2)}(k)$ should be finite except two critical points at the phase transitions between $\mathbb{Z}_{2(3)}$ orders and the disordered phase. As $L$ increases, the peak height of $G^{(2)}(k)$ in the disordered phase saturates to a finite value (first plot of Fig.~\ref{fig:strucfacforonechain}\textbf{c}), while the peak heights at the two critical points keep increasing. Due to the existence of a critical floating phase between $\mathbb{Z}_3$ and $\mathbb{Z}_4$ orders, the peak height of $G^{(2)}(k)$ is larger and increases faster than that in the regime between $\mathbb{Z}_2$ and $\mathbb{Z}_3$ orders (Fig.~\ref{fig:strucfacforonechain}\textbf{c}). For the regime between $\mathbb{Z}_3$ and $\mathbb{Z}_4$ orders with $L=769$, one can see two narrow parts with sharp peaks in $G^{(2)}(k)$ near $\mathbb{Z}_3$ order and $\mathbb{Z}_4$ order, respectively, and the middle part of the regime has relatively smaller peaks. 

To clarify these observations more clearly, we present $G^{(2)}(k)$ as a function of $k$ for five system sizes $L = 25, 97, 241, 385, 769$ and four values of $R_b/a = 2.08, 2.9, 2.97, 3.09$ in Fig.~\ref{fig:strucfacforonechain}\textbf{c}. The system is in the disordered phase for $R_b/a = 2.08, 2.97$ and the peak height $G^{(2)}(k_o)$ at the peak position $k_o$ saturates to a finite value, while the system is in the floating phase for $R_b/a = 2.9, 3.09$ and $G^{(2)}(k_o)$ seems to diverge in the large $L$ limit. Figure~\ref{fig:g2peakscaling1chain} depicts the inverse peak height of $G^{(2)}(k_o)$ as a function of $1/L$ for the 1D Rydberg chain. The data points are fit to the power-law function $a/L^b + g$ and the extrapolated value $g$ is the inverse peak height of $G^{(2)}(k_o)$ in the thermodynamic limit. For $R_b/a=2.04, 2.08,2.12$, the system is in the disordered phase between $\mathbb{Z}_2$ and $\mathbb{Z}_3$ orders. For $R_b/a=2.97$, it is in the disordered phase between $\mathbb{Z}_3$ and $\mathbb{Z}_4$ orders. For $R_b/a=2.9,3.09$, the system is in the floating phase near $\mathbb{Z}_3$ and $\mathbb{Z}_4$ orders, respectively. One can see that the extrapolated value of $1/G^{(2)}(k_o)$ converges to a finite value for the disordered phase, while it approaches zero for the floating phase in the thermodynamic limit. Our analyses thus verify that the connected density-density correlator $G^{(2)}(k_o)$ diverges in the floating phase. All the results presented here are consistent with previous works for the 1D Rydberg chain \cite{PhysRevLett.122.017205, rader2019floating, Chepiga&Mila2021Kibble, PhysRevResearch.3.023049, PhysRevResearch.4.043102}.

\subsection{Two-leg Rydberg Ladder}
We now perform similar analysis for the two-leg Rydberg ladder studied in this work. Figure~\ref{fig:strucfacfortwochains} shows the structure factors $S(k)$ and the connected density-density correlators $G^{(2)}(k)$ for various system sizes. In Fig.~\ref{fig:strucfacfortwochains}\textbf{a}, $S(k)$ as a function of $k$ and $R_b/a \in [1.5, 3.5]$ on four vertical cuts $\Delta/\Omega=2,3,3.5,4$ are presented, where the data sets are the same as those in Fig.~\ref{fig:phasediag}\textbf{b}. The $\Delta/\Omega=2$ cut only goes from $\mathbb{Z}_2$ order into the disordered phase, and thus $S(k)$ only shows sharp peaks at $k=\pi$. Note that the regime where $S(k)$ has prominent peaks with peak positions changing continuously is the floating phase. One can easily identify what density wave orders other cuts go through by looking at the peak of $S(k)$. In particular, the $\Delta/\Omega=3$ cut goes into the disordered phase after $\mathbb{Z}_4$ order, and between $\mathbb{Z}_4$ and $\mathbb{Z}_6$ orders on the $\Delta/\Omega=3.5$ cut is the disordered phase. Notice that between $\mathbb{Z}_2$ and $\mathbb{Z}_3$ orders are the disordered phase followed by a narrow floating phase near $\mathbb{Z}_3$ order, which is supported by the narrow peaks of $S(k)$ near $\mathbb{Z}_3$ plateau on the $\Delta/\Omega=3.5, 4$ cuts (zoom in for better view).

Figure~\ref{fig:strucfacfortwochains}\textbf{b} shows the connected correlator $G^{(2)}(k)$ for the same parameter regime as Fig.~\ref{fig:strucfacfortwochains}\textbf{a}. Similar behaviors for $G^{(2)}(k)$ as those in the 1D Rydberg chain can be seen and only incommensurate regimes between crystalline orders have prominent peaks. The narrow but finite regime of sharp peaks in $G^{(2)}(k)$ before $\mathbb{Z}_3$ regime also suggests a floating phase there. Figure~\ref{fig:strucfacfortwochains}\textbf{c} shows more plots of $G^{(2)}(k)$ at $\Delta/\Omega=4$ and $1.8\le R_b/a \le 3$ for smaller system sizes, where one can see the peak heights of $G^{(2)}(k)$ for the floating phase also increases with increasing $L$. As mentioned before, SBCs discretize the wave vectors for the floating phase and result in steps in the peak plateaus between $\mathbb{Z}_3$ and $\mathbb{Z}_4$ orders. We also show $G^{(2)}(k)$ as a function of $k$ for four system sizes $L=21,33,81,141$ and four sets of parameters $(\Delta/\Omega, R_b/a)$ in Fig.~\ref{fig:strucfacfortwochains}\textbf{d} and plot the finite-size scaling of the inverse peak height of $G^{(2)}(k_o)$ in Fig.~\ref{fig:g2peakscaling2chains}. Similar observations as those in the 1D Rydberg chain can be seen and the same conclusion can be drawn that the connected density-density correlator diverges in the floating phase of the two-leg Rydberg ladder. Importantly, the comparison between the 1D Rydberg chain and the two-leg Rydberg ladder shows that the floating phase region is much broader for the two-leg ladder than the 1D Rydberg chain between $\mathbb{Z}_3$ and $\mathbb{Z}_4$ orders, which greatly facilitates the experimental observation of the floating phase in our experiment using the two-leg ladder setup.

\section{Quantum Phase Transitions in the Two-leg Rydberg Ladder}
Since our experiments can probe both $\mathbb{Z}_p$ crystalline orders and the floating phase, the associated quantum phase transitions can be studied in future works. Here, we provide numerical studies for the quantum phase transitions in the two-leg Rydberg ladder. Previous works studied the quantum phase transitions in the 1D Rydberg chain by looking at divergence of the correlation length extracted from DMRG calculations of correlation functions for very large systems (up to $L = 9000$) \cite{PhysRevLett.122.017205,Chepiga&Mila2021Kibble,PhysRevResearch.3.023049,PhysRevResearch.4.043102}. Here, we use Binder Cumulant $U_4$ and the energy gap $\Delta E$ to locate the phase transition point and determine the correlation length exponent $\nu$ and the dynamical exponent $z$. The Binder Cumulant $U_4$ is defined by
\begin{eqnarray}
    \label{eq:binderc}
    U_4 = 1 - \frac{\langle M_k^4 \rangle}{3\langle M_k^2 \rangle^2},
\end{eqnarray}
where $M_k$ is the order parameter of the density wave order with wave vector $k$. If $R_b/a$ is fixed, at the phase transition point between two gapped phases, the correlation length diverges as $\xi \sim \left|\Delta/\Omega - (\Delta/\Omega)_c \right|^{-\nu}$, where $(\Delta/\Omega)_c$ is the value of $(\Delta/\Omega)$ at the critical point and $\nu$ is the correlation length exponent. The Binder Cumulant satisfies the functional form 
\begin{eqnarray}
U_4 = f \left[L^{1/\nu} \left(\Delta/\Omega-(\Delta/\Omega)_c \right) \right]
\end{eqnarray}
up to a subleading correction around the critical point, where $f(x)$ is some universal function. It is a dimensionless quantity and has a fixed point at the phase transition point. In the large $L$ limit, $U_4$ with different $L$s should intersect at the phase transition point $(\Delta/\Omega)_c$. By finding the best data collapse of $U_4$ for different system sizes, we can determine the values of both $(\Delta/\Omega)_c$ and $\nu$.

The dynamical exponent $z$ is defined such that the dispersion relation $\omega \sim k^z$ for small $k$ at the critical point. This means that the characteristic time of the system can relate to the correlation length by $\tau_c \sim \xi^z$ \cite{HohenbergDynamic1977}. For phase transition points that are described by CFTs, there is a linear dispersion relation and $z = 1$. There also exist other phase transition points like chiral transitions with $z > 1$ \cite{CARDY1993577,NYCKEES2021115365}. Numerically, the energy gap $\Delta E$ between the first excited state and the ground state in finite systems can be used to determine $z$, utilizing the ansatz
\begin{eqnarray}
L^z\Delta E = g \left[L^{1/\nu} \left( \Delta/\Omega - (\Delta/\Omega)_c \right) \right],
\end{eqnarray}
where $g$ is another universal function. The best data collapse of the rescaled energy gap $L^z\Delta E$ can simultaneously give values of $(\Delta/\Omega)_c$, $\nu$, and $z$. In practice, we first fit $U_4$ as a function of $L^{1/\nu}[\Delta/\Omega-(\Delta/\Omega)_c]$ in a small window around the critical point to a high-degree polynomial, scanning the values of $(\Delta/\Omega)_c$ and $\nu$ to minimize the residual sum of squares. Then, we perform the same procedure for the energy gap with fixed $(\Delta/\Omega)_c$ and $\nu$ obtained from $U_4$ to determine $z$.

In the following subsections, we carefully investigate the quantum phase transitions across the $\mathbb{Z}_2, \mathbb{Z}_3$, and $\mathbb{Z}_4$ lobes. Through thorough numerical analysis, we show the existence of a variety of quantum phase transitions, including Ising phase transitions, conformal and chiral phase transitions, and BKT and PT phase transitions, which are similarly present in the 1D Rydberg chain~\cite{PhysRevLett.122.017205, rader2019floating, Chepiga&Mila2021Kibble, PhysRevResearch.3.023049, PhysRevResearch.4.043102}. Compared to the 1D chain, however, some of the phase transitions are more readily accessible in experiment in the two-leg Rydberg ladder setup. Hence, our theoretical analysis here paves the way for near-future experimental studies of rich quantum phase transition physics experimentally accessible in the two-leg Rydberg ladder.

\begin{figure*}[t]
\centering
\includegraphics[width=1\textwidth]{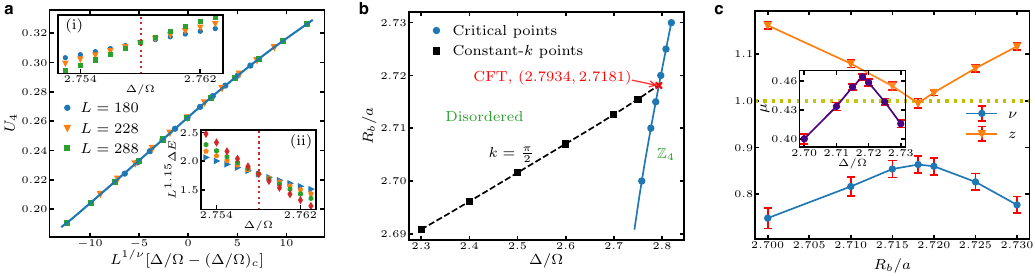}
    \caption{\textbf{Chiral and CFT phase transitions between the $\mathbb{Z}_4$ order and the disordered phase.} \textbf{a}, Collapse of Binder Cumulant $U_4$ as a function of $L^{1/\nu}[\Delta/\Omega-(\Delta/\Omega)_c]$ around the quantum phase transition along the $R_b/a=2.7$ cut. Inset (i) shows the same set of $U_4$ as a function of the original parameter $\Delta/\Omega$. Inset (ii) shows the rescaled energy gap $L^z\Delta E$ as a function of $\Delta/\Omega$ for $L=96,120,144,180$ (clockwise). \textbf{b}, The CFT point on the phase boundary of $\mathbb{Z}_4$ order is at the point of intersection between the $k=\pi/2$ commensurate line and the phase boundary. This is labeled as the white circle in Fig.~\ref{fig:entphasediag} on the $\mathbb{Z}_4$ phase boundary. \textbf{c}, The critical exponents $\nu$ and $z$ as functions of $R_b/a$ for chiral and CFT transitions on the boundary of $\mathbb{Z}_4$ order. The inset shows the Kibble-Zurek exponent $\mu$ as a function of $R_b/a$.}
    \label{fig:chiralcftz4}
\end{figure*}
\begin{figure}
\centering
\includegraphics[width=\linewidth]{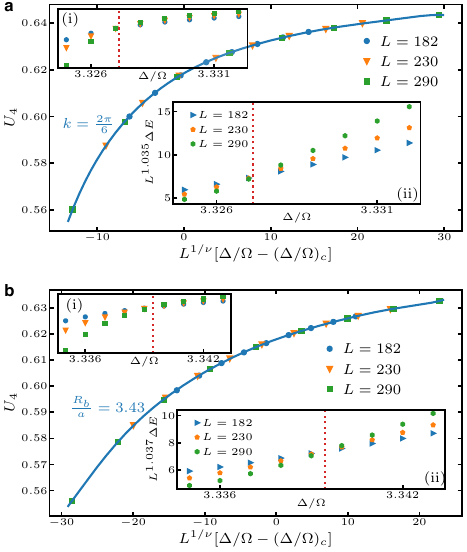}
    \caption{\textbf{Quantum phase transitions between the $\mathbb{Z}_6$ order and the disordered phase.} The plots are the same as Fig.~\ref{fig:chiralcftz4}\textbf{a}, but for the phase transitions along the commensurate line $k=2\pi/6$ (\textbf{a}), which is the CFT point labeled as a white circle in Fig.~\ref{fig:entphasediag}, and the horizontal cut $R_b/a=3.43$ (\textbf{b}).}
    \label{fig:chiralcftz6}
\end{figure}
\begin{figure*}
\centering
\includegraphics[width=1\textwidth]{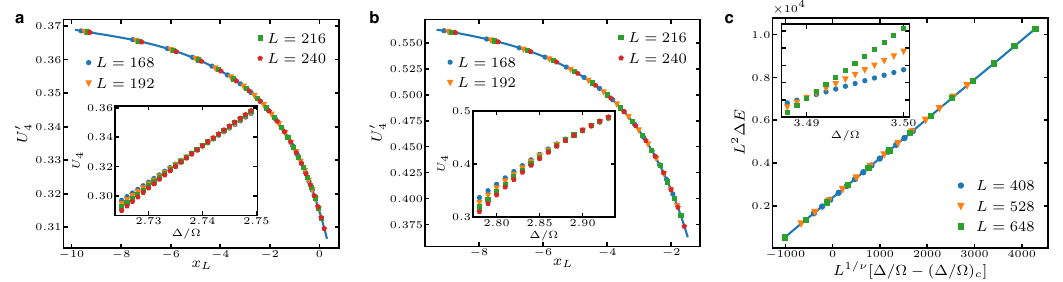}
    \caption{\textbf{BKT transitions between the floating phase and the disordered phase, and PT transitions between the floating phase and the $\mathbb{Z}_3$ order.} \textbf{a}, Collapse of the modified Binder Cumulant $U'_4$ for the BKT transition along the commensurate line $k=2\pi/3$. The inset shows $U_4$ as a function of $\Delta/\Omega$ for the same system sizes. \textbf{b}, Same as \textbf{a}, but for the incommensurate line $k=2\pi \times 7/24$. \textbf{c}, Collapse of the rescaled energy gap $L^2\Delta E$ inside the $\mathbb{Z}_3$ order but near the PT transition along the horizontal cut $R_b/a = 2.3$. The inset shows $L^2\Delta E$ as a function of $\Delta/\Omega$.}
    \label{fig:bktpttransitions}
\end{figure*}

\subsection{Ising Phase Transitions}
The domain wall excitations on top of the $\mathbb{Z}_2$ symmetry breaking state have only one type and no chiral perturbation is introduced. As shown in Fig.~\ref{fig:phasediag}(b), the peak position of the structure factor around $\mathbb{Z}_2$ order is constant at $p=2$ and $k=\pi$. The quantum phase transitions between $\mathbb{Z}_2$ order and disordered phases are of Ising CFT points with $\nu=1$ and $z=1$. To show the numerical evidence and test the validity of data collapse, we present Binder Cumulant and rescaled energy gap for four phase transition points on the boundary of $\mathbb{Z}_2$ order in Fig.~\ref{fig:collapseu4gapising}.

We note that the regime of $\mathbb{Z}_2$ order is at $1< R_b/a \lesssim 2$, where NN sites in the same leg are blockaded but the two sites in the same rung are not. The local degrees of freedom that do not favor $\mathbb{Z}_2$ order can introduce large finite-size effects. We perform data collapse of $U_4$ for large system sizes $L=408, 528, 648$ and that of $L^z \Delta E$ for $L=288,348,408$ to reduce the finite-$L$ effects. The first phase transition point we studied is the intersection between the $\Delta/\Omega=3.3$ cut and the lower boundary of $\mathbb{Z}_2$ order. We calculate $U_4$ and $\Delta E$ at nine values of $R_b/a$ starting from $1.1474$ with an increment of $0.0001$. The best data collapse is found at $(R_b/a)_c = 1.14790$, $\nu=0.986$. One can see that all the data for $U_4$ with different $L$s collapse onto a smooth curve that is a function of $L^{1/\nu}[R_b/a - (R_b/a)_c]$. The inset in the first plot of Fig.~\ref{fig:collapseu4gapising}\textbf{a} shows $U_4$ versus $R_b/a$ for different $L$s, which cross exactly at $(R_b/a)_c$. Fixing $(R_b/a)_c=1.14790$ and $\nu=0.986$, we scan values of $z$ and the best data collapse for $L^z\Delta E$ happens at $z=1.01$. The inset in the first plot of Fig.~\ref{fig:collapseu4gapising}\textbf{b} shows $L^z\Delta E$ versus $R_b/a$, which also cross exactly at $(R_b/a)_c$. The same procedures are performed on the other three phase transition points on $R_b/a = 1.3, 1.5, 2.05$ cuts, respectively. The data points take nine values of $\Delta/\Omega$ starting from $1.034$ with an increment of $0.001$, nine values of $\Delta/\Omega$ starting from $1.124$ with an increment of $0.001$, and eleven values of $\Delta/\Omega$ starting from $4.0848$ with an increment of $0.0001$, respectively. The phase transition points and critical exponents are found to be $((\Delta/\Omega)_c, \nu, z)=(1.03851, 0.983, 1.03)$, $(1.12870, 0.981, 1.02)$, and $(4.08531, 0.978, 1.03)$, respectively. A smaller truncation error $\epsilon=10^{-11}$ is used for the $R_b/a=2.05$ cut, because the phase transition is very close to other critical regime (the floating phase below $\mathbb{Z}_3$ order). One can see that all the values of $\nu$ and $z$ obtained from data collapse are consistent with $\nu=1$ and $z=1$ for Ising transitions. Data collapse of Binder Cumulant and the energy gap is a reliable method to extract critical points and critical exponents in Rydberg systems.

\subsection{Conformal and Chiral Phase Transitions}

We now focus on the conformal and chiral quantum phase transitions between the $\mathbb{Z}_{4(6)}$ order and the disordered phase. As mentioned above, on the boundary of $\mathbb{Z}_{4(6)}$ order, there exists a segment of direct transitions between the crystalline order and the disordered phase. The segment contains a CFT point with $z=1$ and other points are chiral transition points with $z$ continuously changing from $1$ to $2$. We again use data collapse of $U_4$ and $L^z\Delta E$ to determine the phase transition points and critical exponents $\nu$ and $z$ for chiral phase transitions. One example on data collapse for the chiral transition along $R_b/a=2.7$ cut is shown in Fig.~\ref{fig:chiralcftz4}\textbf{a}. We calculate $U_4$ and $\Delta E$ for eleven values of $\Delta/\Omega$ taking from $2.753$ with an increment of $0.001$. Three system sizes $L=180,228,288$ are used for the collapse of $U_4$ and four system sizes $L=96,120,144,180$ are used for the collapse of $L^z\Delta E$. The best data collapse is found at $(\Delta/\Omega)_c = 2.75805$, $\nu = 0.726$, and $z=1.15$. The inset (i) [(ii)] shows that $U_4$ ($L^z\Delta E$) as a function of $\Delta/\Omega$ for different system sizes cross exactly at the critical point $(\Delta/\Omega)_c$. We use the same method to find the critical points and critical exponents for other $R_b/a$ cuts. The critical points are plotted in Fig.~\ref{fig:chiralcftz4}\textbf{b} and cubic spline interpolation is applied to draw the boundary of $\mathbb{Z}_4$ order.

One can calculate more critical points to precisely locate the CFT point with $z=1$. Meanwhile, the CFT point is the intersection between the $\mathbb{Z}_4$ boundary and the commensurate line $k=\pi/2$ where the chiral perturbation from domain-wall excitations is zero. From the phase diagram in Fig.~\ref{fig:phasediag}\textbf{b}, it is found that the commensurate lines $k=2\pi/3, 2\pi/4, 2\pi/6$ meet the crystalline lobes around $R_b/a = 2.12, 2.71, 3.44$, respectively (also shown in Fig.~\ref{fig:entphasediag}). To accurately determine the CFT points, we precisely locate the peak position of the structure factor $S(k)$ for various values of $R_b/a$ with $0.001$ step on constant $\Delta/\Omega$ cuts and use spline interpolation to find the commensurate point with $k=\pi/2$ for $L=144,180,228,288,348$. Then, we extrapolate the commensurate points to the values in the thermodynamic limit for each constant $\Delta/\Omega$ cut. The extrapolated commensurate points are plotted in Fig.~\ref{fig:chiralcftz4}\textbf{b}. Another extrapolation procedure is performed to find that the commensurate line intersects with the $\mathbb{Z}_4$ boundary at the CFT point $(\Delta/\Omega, R_b/a)_{\rm{CFT}} = (2.7934, 2.7181)$. To double check that the intersection is a CFT point, we perform data collapse on $R_b/a=2.7181$ cut: the critical point and exponents are found to be $[(\Delta/\Omega)_c, \nu, z] = (2.79343, 0.8407, 0.99)$, where the value of $z$ is consistent with that the transition point is conformal with $z=1$. The value of $\nu$ obtained here is consistent with the predicted regime $\nu \in (0.683, 1]$ where the chiral perturbation is relevant for the Ashkin-Teller model describing the universality classes for physical systems with four effective degrees of freedom \cite{AshkinTeller1943,Kohmoto1981AT,schulz1983phase}.

We further calculate the results for the CFT point with smaller truncation error $\epsilon=10^{-11}$. It is found that $[(\Delta/\Omega)_c, \nu, z] = (2.79356, 0.8636, 0.995)$ for the CFT point. The results are very close to those for $\epsilon=10^{-10}$ and the value of $z$ increases a little towards the expected value $z=1$. Based on this, one can estimate that the errors of the critical points, $\nu$, and $z$ calculated with $\epsilon=10^{-10}$ is of order $10^{-4}$, $10^{-2}$, and $10^{-2}$, respectively. We can also correct the values of $\nu$ and $z$ for other critical points calculated with $\epsilon=10^{-10}$ based on the changes for the CFT point with smaller $\epsilon=10^{-11}$, as long as the critical points are not far from the CFT point. We plot values of $\nu$ and $z$ as functions of $R_b/a$ in Fig.~\ref{fig:chiralcftz4}\textbf{c}. It is seen that $\nu$ reaches maximum at the CFT point, while $z$ gets to the minimum about $1$ at the CFT point. The results clearly verify that the phase transitions between the $\mathbb{Z}_4$ order and the disordered phase is chiral with $z>1$ except at a single CFT point with $z=1$. The Kibble Zurek exponent $\mu = \nu / (1+z\nu)$ \cite{TWBKibble_1976,Zurek1985CosHelium,Polkovnikov2005dynamics,ZurekDorner2005,DziarmagaDynamicsIsing2005} is calculated and shown in the inset of Fig.~\ref{fig:chiralcftz4}\textbf{c}, where we see that the maximum $\mu_{\rm{max}} = 0.465(3)$ is also at the CFT point. The critical exponents $\nu$ and $z$ will continuously decrease and increase respectively until $\nu=1/2$ and $z=2$, where the chiral transition line ends at Lifshitz points and splits into one BKT line and one PT line that surround a critical floating phase \cite{PhysRevResearch.3.023049}. 

Unlike the 1D Rydberg chain where $\mathbb{Z}_6$ order cannot directly melts into the disordered phase, the two-leg Rydberg ladder considered here has a direct transition line between the $\mathbb{Z}_6$ order and the disordered phase. In Fig.~\ref{fig:chiralcftz6}, we study the data collapse on the commensurate line $k=2\pi/6$ and the $R_b/a=3.43$ cut. Here, the truncation error in DMRG $\epsilon=10^{-12}$ is used for good accuracy. The commensurate line $k=2\pi/6$ is determined by the same way as that for $k=2\pi/4$ line in Fig.~\ref{fig:chiralcftz4}\textbf{b}. Notice that $L\equiv 2 \pmod{6}$ is required for the ladder system with SBCs to be compatible with the $\mathbb{Z}_6$ order, so we use $L=182, 230, 290$ here for the data collapse. The critical points and critical exponents are found to be $[(\Delta/\Omega)_c, (R_b/a)_c, \nu, z] = (3.32713, 3.422, 0.651, 1.035)$. The value of $z$ is consistent with the expected value $z=1$ for CFT. Thus, the quantum phase transition along the $k=2\pi/6$ line is also a CFT point. On $R_b/a=3.43$ cut, the critical point is close to the CFT point on the commensurate line. It is expected that $z$ is still close to $1$. As shown in Fig.~\ref{fig:chiralcftz6}\textbf{b}, $[(\Delta/\Omega)_c, \nu, z] = (3.33945, 0.647, 1.037)$, which is consistent with our expectation. More chiral critical points and critical exponents on $\mathbb{Z}_4$ and $\mathbb{Z}_6$ lobes are beyond the scope of this work and will be left for future work.

\subsection{BKT and PT Phase Transitions}
In this section, we numerically demonstrate the existence of BKT and PT transitions. As mentioned before, the floating phase is a critical phase with the central charge $c=1$. Thus, the dispersion relation is linear and the low-energy effective theory is a TLL. TLL depends on two parameters: the Luttinger liquid parameter $K$ and the velocity $v$ \cite{F.D.M.Haldane_1981,J_Voit_1995}. If $K$ exceeds some critical value, TLL becomes unstable and a BKT transition to the disordered phase happens. Going towards BKT transitions from the gapped side, the correlation length diverges fast and approaches an essential singularity: $\xi \sim \exp(b/\sqrt{\delta})$ \cite{Kosterlitz_1974}, where $b$ is a constant and $\delta$ is the distance to the BKT point. If the velocity $v=0$, TLL does not flow and a PT transition to crystalline orders happens, where the linear dispersion vanishes and the quadratic dispersion dominates. Approaching PT transitions from the gapped side, the correlation length diverges as $\xi \sim \delta^{-1/2}$ and the energy gap $\Delta E \sim 1/L^2$; that is, the critical exponents $\nu = 1/2$ and $z = 2$.

The Binder Cumulant can also be used to probe BKT transitions. In this case, $U_4$ is a universal function of $x_L = \ln(L/\xi) = \ln L - b/\sqrt{\delta}$. As logarithmic corrections exist for BKT transitions, a correction factor is applied to increase accuracy $U_4 \rightarrow U'_4 = (1+A/\ln L)U_4$, where $A$ is a constant. Therefore, collapse of Binder Cumulant for BKT transitions takes the form $U'_4 = f(x_L)$. We probe BKT transitions along one commensurate line $k=2\pi/3$ and one incommensurate line $k = 2\pi\times 7/24$. The two constant-$k$ lines are found by the extrapolation method described before. We have mentioned that there is no direct melting of the $\mathbb{Z}_3$ order into the disordered phase and a very narrow floating phase below the $\mathbb{Z}_3$ lobe exists. To verify that, we only need to show that there exists a BKT transition along the commensurate line $k=2\pi/3$ because the chiral perturbations in incommensurate regime will further favor the existence of the floating phase \cite{NYCKEES2021115365,CoppersmithDisloc1982}. We first assume that the $\mathbb{Z}_3$ order can directly melt into the disordered phase and the Binder Cumulant still satisfies the form $U_4 = f[L^{1/\nu}(\Delta/\Omega-(\Delta/\Omega)_c)]$. However, an unreasonably large $\nu \approx 3$ is found using data for $L \le 240$. The extracted value of $\nu$ increases quickly by using more data with larger system sizes. These observations reveal the existence of an infinite-order BKT transition and there is no direct transition between the $\mathbb{Z}_3$ order and the disordered phase. We then use the BKT form $U'_4 = f(x_L)$ for data collapse as shown in Fig.~\ref{fig:bktpttransitions}. The insets of Fig.~\ref{fig:bktpttransitions}\textbf{a} and Fig.~\ref{fig:bktpttransitions}\textbf{b} show that $U_4$ for different $L$s nearly coalesce after the intersection, indicating a gapped-to-gapless transition. The best data collapse is found at $(\Delta/\Omega, R_b/a)_{\rm{BKT}} = [2.747(5), 2.1049(2)]$ for $k=2\pi/3$ and $[2.976(5), 2.4159(3)]$ for $k=2\pi\times 7/24$. The values of $b$ are determined as $0.78(17)$ and $3.05(14)$, respectively.

We finally use collapse of the rescaled energy gap to determine the values of $\nu$ and $z$ for PT transitions. The ansatz $L^z\Delta E = f[L^{1/\nu}(\Delta/\Omega-(\Delta/\Omega)_c)]$ is the same as the one used before. As there is no collapse of $U_4$ near PT transitions, we just use the collapse of $L^z\Delta E$ to determine the critical points and critical exponents simultaneously. The results for the $R_b/a = 2.3$ cut are shown in Fig.~\ref{fig:bktpttransitions}\textbf{c}, where data for $L=408,528,648$ with $\epsilon=10^{-11}$ are used. The best data collapse is found at $[(\Delta/\Omega)_c, \nu, z] = (3.4908, 0.55, 1.80)$. If we fix $z = 2$, the best data collapse is found at $(\Delta/\Omega)_c = 3.4903$ and $\nu = 0.4981$. All these results are consistent with theoretical predictions $\nu=1/2$, $z=2$ for PT transitions. Therefore, the phase transition between the floating phase and the crystalline orders is of PT universality class.

\end{document}